\documentclass[sigconf]{acmart}
\AtBeginDocument{%
  }
\newcommand{\modelname}{TextBridgeGNN}

\usepackage{multirow}
\usepackage{colortbl}
\usepackage{xcolor}
\usepackage{booktabs}
\usepackage{graphicx}
\usepackage{subcaption} 
\usepackage{hhline}
\usepackage{bm} 
\usepackage{siunitx} 
\usepackage{amsmath} 
\usepackage{tikz}
\usepackage{tabularx}
\usepackage{dashrule}

\begin{document}

\title{TextBridgeGNN: Pre-training Graph Neural Network for Cross-Domain Recommendation via Text-Guided Transfer}

\author{Yiwen Chen}
\authornote{Both authors contributed equally to this research.}
\email{yiwenchen@buaa.edu.cn}
\affiliation{%
  \institution{Beihang University}
  \city{Beijing}
  \country{China}
}

\author{Yiqing Wu}
\authornotemark[1]
\email{iwu_yiqing@163.com}
\affiliation{%
  \institution{Institute of Computing Technology, Chinese Academy of Sciences}
  \city{Beijing}
  \country{China}
}

\author{Huishi Luo}
\email{hsluo2000@buaa.edu.cn}
\affiliation{%
  \institution{Beihang University}
  \city{Beijing}
  \country{China}
}

\author{Fuzhen Zhuang}
\authornote{Corresponding author.}
\email{zhuangfuzhen@buaa.edu.cn}
\affiliation{%
  \institution{Institute of Artificial Intelligence, Beihang University}
  \city{Beijing}
  \country{China}
}

\author{Deqing Wang}
\email{dqwang@buaa.edu.cn}
\affiliation{%
  \institution{Beihang University}
  \city{Beijing}
  \country{China}
}

\author{Zhao Zhang}
\email{zhao_zhang@buaa.edu.cn}
\affiliation{%
  \institution{Beihang University}
  \city{Beijing}
  \country{China}
}

\renewcommand{\shortauthors}{Y. Chen et al.}


\begin{abstract}
Graph-based recommendation has achieved great success in recent years. The classical graph recommendation model utilizes ID embedding to store essential collaborative information. However, this ID-based paradigm faces challenges in transferring to a new domain. This phenomenon primarily stems from two inherent challenges: (1) the non-transferability of ID embeddings due to isolated domain-specific ID spaces, and (2) structural incompatibility between heterogeneous interaction graphs across domains. To address these issues, we propose \modelname{}, a pre-training and fine-tuning framework that can effectively transfer knowledge from a pre-trained GNN to downstream tasks. Specifically, \modelname{} uses text as a semantic bridge to connect domains through multi-level graph propagation. During the pre-training stage, hierarchical GNNs are designed to learn both domain-specific and domain-global knowledge with text features, ensuring the retention of collaborative signals and the enhancement of semantics. During the fine-tuning stage, a similarity transfer mechanism initializes ID embeddings in the target domain by transferring from semantically related nodes, successfully transferring the ID embeddings and graph pattern. Experiments demonstrate that \modelname{} consistently achieves strong overall performance across cross-domain, multi-domain, and training-free scenarios, without costly language model fine-tuning or real-time inference overhead.
\end{abstract}

\vspace{-10pt}

\ccsdesc[500]{Information systems~Recommender systems\vspace{-5pt}}

\keywords{\vspace{-2pt}Cross-Domain Recommendation, Graph Neural Networks (GNNs)\vspace{-2pt}}

\maketitle
\vspace{-12pt}
\section{Introduction}

Recommender systems aim to provide appropriate items by analyzing user preferences contained in the user behaviors. Generally, the interactions between users and items naturally form a graph structure. Inspired by the success of graph neural networks (GNNs), the graph-based recommendation achieves great success\cite{classicgnn:conf/iclr/VelickovicCCRLB18,classicgnn:journals/corr/KipfW16,classicgnn:journals/tnn/WuPCLZY21,classichnn:conf/nips/HamiltonYL17}. Most of these are ID-based graph recommendation,
which assign unique ID embeddings to each user and item and model the interaction relationships using the GNNs. Existing graph and non-graph recommendation models have shown that ID embeddings effectively capture collaborative filtering signals and have become a core component of recommender systems. However, when facing the demand of cross-domain recommendation or pre-training, ID-based graph recommendation models encounter two fundamental issues: the non-transferability of ID embeddings (due to independent ID spaces in different domains leading to knowledge fragmentation) and the domain differences in graph structures (heterogeneous interaction graph topologies hindering generalization capabilities).

To address the above issues, numerous efforts have already been undertaken.
The first category of methods attempts to transfer knowledge by overlapping users/items\cite{M2GNN:conf/sigir/HuaiYZZL023, 13DBLP:conf/cikm/LiuLLP20, 32:conf/ijcai/ZhuWCLZ20, 14Cross-GraphKnowledgeTransferNetwork:conf/icc/00030W021, IIGCN:conf/www/HanZCCY23, RLCDR:journals/tkde/LiHL24}. They utilize the overlapping users/items to merge the source domain and target domain into one graph. For example,
CGKT \cite{14Cross-GraphKnowledgeTransferNetwork:conf/icc/00030W021} designs different aggregation functions to model cross-domain relationships and single-domain relationships.
EDDA\cite{EDDA:conf/cikm/Ning0LCZT23} further splits ID embeddings into general and domain-specific components.
However, in the real world, it is challenging to find overlapping users and items across domains, limiting the universality of these methods.

The second category of methods attempts to bypass ID embeddings and turn to text-driven cross-domain transfer.
As the textual information and word tokens are universal across domains. Those methods replace IDs with textual information and utilize the generalization capabilities of pre-trained language models to transfer knowledge.  While those methods still face several issues: 1) Some cross-domain recommendation works have tried using pre-trained language models (e.g., UniSRec\cite{Unisrec:conf/kdd/HouMZLDW22}, VQRec\cite{VQrec:conf/www/HouHMZ23}) to construct universal representations based on the text features of user behavior sequences. \textbf{However, text features struggle to replace the collaborative signals implicit in ID embeddings (such as implicit group preferences).} 2) Although some graph recommendation methods, such as MMGCN\cite{MMGCN:conf/mm/WeiWHHC19} and LLMRec\cite{LLMRec:conf/wsdm/WeiRTWSCWYH24}, attempt to combine ID and text features to consider both semantic information in textual features and collaborative information in ID embeddings.   They are not specifically designed for cross-domain recommendation, \textbf{still lacking core methods for ID embedding transfer.} 3) Some LLM-driven methods (e.g., Uni-CTR\cite{UniCTR:journals/corr/abs-2312-10743}) require fine-tuning large parameters to adapt to recommendation tasks, with computational overhead exceeding the real-time requirements of industrial systems.
\textbf{These three issues expose the essential contradiction of the text-driven paradigm: sacrificing the transfer of core collaborative signals and graph structure knowledge in recommendation in exchange for partial generalization capabilities.}

Despite the success of previous cross-domain/pre-trained recommendation models, few studies have explored applying pre-training and fine-tuning paradigms to graph-based cross-domain recommendation. Although Wang et al. \cite{20pretrainGraph_wang2021pre} proposed a contrastive pre-training framework to alleviate structure bias, its effectiveness is limited by a partial parameter transfer that excludes ID embeddings and an MF-based fine-tuning stage that is misaligned with the graph pre-training. Other works\cite{19:conf/kdd/QiuCDZYDWT20, GPPT:conf/kdd/SunZHWW22, gpt-gnn:conf/kdd/HuDWCS20} are not specifically optimized for cross-domain recommendation, leaving challenges like non-transferable ID embeddings and domain-specific graph structures largely unaddressed. Recently, AlphaRec\cite{AlphaRec25} has shown that graph-based recommendation can achieve good domain generalization using only item text representations. However, it bypasses the challenge of transferring ID-based collaborative signals, which remains an open question.

\begin{figure}[t]
    \centering
    \includegraphics[width=\columnwidth]{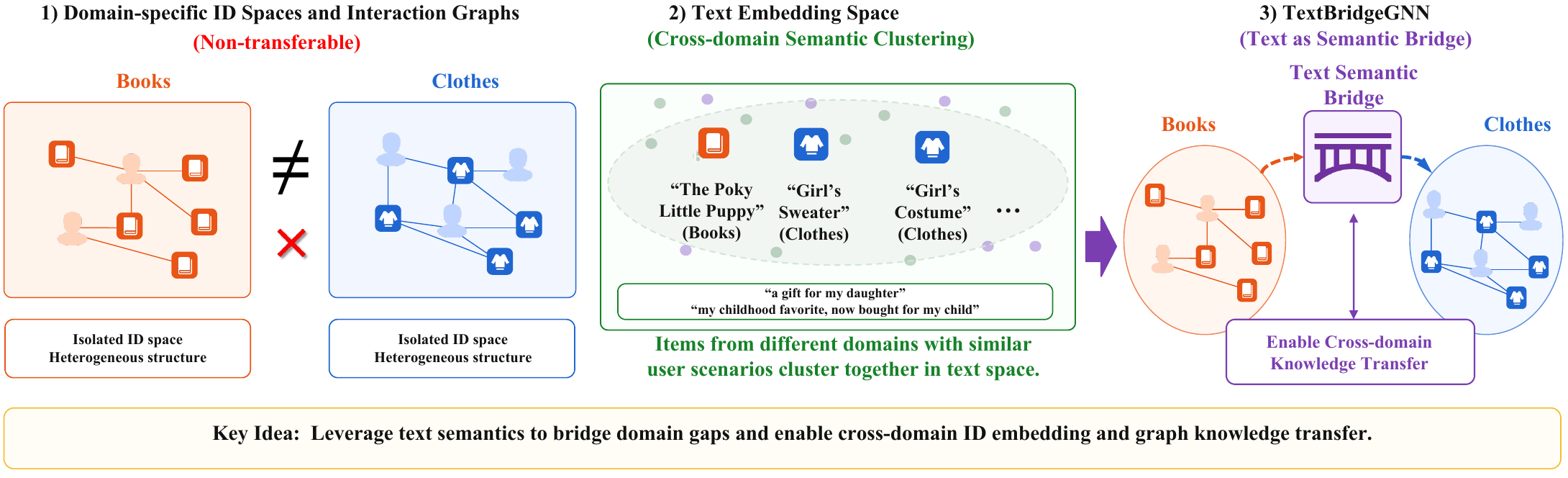}
    \caption{Motivation of \modelname{}.}
    \label{fig:motivation}
    \vspace{-2pt}
\end{figure}

To address those issues, we start by considering what needs to be transferred in pre-training for graph recommendation. Similar to other domains, the most obvious step is to transfer the pre-trained parameters. In the case of recommendation, this includes the parameters of the GNN and the ID embeddings. However, unlike other domains, structural information is also critical in graphs. An intuitive approach is to directly utilize the original graph data, as graph structural information is typically captured through the message-passing mechanism of GNNs, which conveys the k-hop neighborhood information to a node. The key insight is that \textbf{text is not merely an auxiliary feature but a necessary bridge}: since ID embeddings are inherently non-transferable across domains and graph structures are incompatible, text-derived semantic similarity is the only viable medium to establish cross-domain alignment while preserving collaborative signals.

Based on this, this paper proposes the \modelname{}, whose core idea is to \textbf{align cross-domain knowledge using text as a semantic bridge while retaining the collaborative signals of ID embeddings and the high-order associations of graph structures}. During the pre-training phase, we designed a hierarchical pre-training mechanism that constructs domain-specific subgraphs and a global graph to simultaneously learn domain-specific and domain-general knowledge.
In the fine-tuning phase, we addressed the ID mismatch problem by establishing semantic edges between the upstream and downstream graphs using textual information. A hierarchical graph similarity transfer module helps the model transfer both domain-specific and domain-general knowledge effectively. Overall, our contributions can be summarized as follows:
\vspace{-3pt}
\begin{itemize}
    \item To the best of our knowledge, we are the first to propose an ID-based graph pre-training recommendation model in a universal setting that incorporates textual information as a bridge for domain transfer. Our \modelname{} can effectively transfer the collaborative information embedded in ID embeddings, as well as the graph structural information across multiple domains during pre-training.
    \item We designed a hierarchical knowledge learning mechanism incorporating semantic information that enables the model to simultaneously learn domain-specific knowledge and multi-domain global knowledge during the pre-training phase. During the fine-tuning phase, it can effectively transfer this knowledge separately through textual information.
    \item We conduct extensive experiments on two real-world datasets. The results demonstrate that \modelname{} could effectively transfer knowledge from pre-trained graph models. And \modelname{} achieves strong performance across multiple scenarios.
\end{itemize}
\vspace{-6pt}

\begin{figure*}[htbp]
    \centering
    \includegraphics[width=\textwidth]{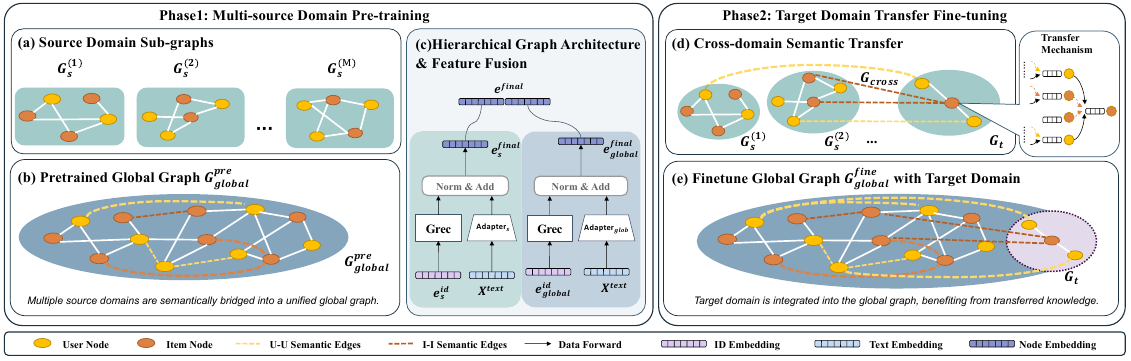}
    \caption{Model Architecture: \modelname{} consists of two main phases: (1) Multi-domain pre-training, which employs a hierarchical message passing mechanism to capture both local and global interactions within and across domains while fusing text and ID embeddings; (2) Cross-domain fine-tuning, which uses a hierarchical graph similarity transfer framework to transfer knowledge from the source domain to the target domain. }
    \label{fig:model_architecture}
    \vspace{-6pt}
\end{figure*}

\section{Methodology}

\subsection{Problem Formulation}
In this paper, we focus on ID-based graph model pre-training in recommendation, which generally involves users $u \in U$, items $v \in V$, and an interaction graph $\mathcal{G} = (\mathcal{U}, \mathcal{V}, \mathcal{E})$, where $\mathcal{E}$ is the set of interaction edges between $u$ and $v$.  The graph pre-training recommendation aims to pre-train a graph neural network (GNN) on multi-domains, and this model can be applied to downstream domains. Given graph data from $N$ domains $\{\mathcal{G}_s^{(i)}\}_{i=1}^N$, we first train a pre-trained GNN recommendation model $f_{g}(\{\mathcal{G}_s^{(i)}\}_{i=1}^N|\theta)$, where $\theta$ is the pre-trained parameters. After that, the pre-trained model can be directly applied or fine-tuned on the downstream domain interaction graph  $\mathcal{G}_t = (\mathcal{U}_t, \mathcal{V}_t, \mathcal{E}_t)$ for recommendation.
\vspace{-9pt}

\subsection{Overall Framework}
In this paper, we propose \modelname{} to address graph pre-training in recommendation.  As shown in Figure.\ref{fig:model_architecture}, similar to most pre-training pipelines, our \modelname{} includes two phases: (1) multi-source domains pre-training and (2) downstream fine-tuning. In the multi-source domains pre-training stage, we carefully design a
 \textbf{hierarchical message passing mechanism}, which is capable of comprehensively considering both the distinctive features of individual domains and the interconnections among overall domains within the multi-domain pre-training data. In the fine-tuning stage, we first utilize the textual information as a bridge to connect the pre-training graph and the downstream graph, thus transferring the graph structure information and collaborative information in the ID embedding. Then, the pre-trained model is fine-tuned on the downstream data for downstream domain recommendation.

We introduce text feature matrices for users and items, denoted as $\mathbf{X}_u^{text} \in \mathbb{R}^{|\mathcal{U}| \times d_{text}}$ and $\mathbf{X}_v^{text} \in \mathbb{R}^{|\mathcal{V}| \times d_{text}}$, where $|\mathcal{U}|$ and $|\mathcal{V}|$ represent the total number of users and items across all domains, and $d_{text}$ denotes the dimensionality of the text features. These text features capture semantic information of users and items, providing support for cross-domain transfer.

The recommendation function in target domain is defined as:

    \vspace{-5pt}
\begin{small}
    \begin{equation}
        f_t(u,v) = \sigma\left(\mathbf{h}_u^{final} \cdot \mathbf{h}_v^{final}\right),
    \end{equation}
\end{small}

    \vspace{-5pt}

where $\sigma(\cdot)$ is the Sigmoid function, and $\mathbf{h}_u^{final}$ and $\mathbf{h}_v^{final}$ represent the final embeddings of users and items, respectively.

\vspace{-1pt}

As shown in Figure.\ref{fig:prompt_pipeline}, we generate semantic embeddings $\mathbf{X}_u^{text} \in \mathbb{R}^{|\mathcal{U}| \times d_{text}}$ and $\mathbf{X}_v^{text} \in \mathbb{R}^{|\mathcal{V}| \times d_{text}}$ by leveraging large language models (LLMs) to encode textual and interaction-based information into dense vector representations. The pipeline consists of preprocessing textual data, summarizing interaction histories, generating prompts, and finally encoding them into semantic embeddings using models such as \texttt{Llama 3}\cite{llama3modelcard}. (We provide more details in Appendix.\ref{appendix:prompt_details} and ablation study in section \ref{subsec:robustness})



\subsection{Multi-source Domain Graph Pre-training Phase}
Generally, pre-trained language models are trained on multi-domains, thereby, models can integrate knowledge in diverse domains and be more generalized.  However,  in the recommender system, things become different. Unlike tokens that are universal in NLP, different domains (such as e-commerce, short videos, and news recommendation) often adopt entirely independent ID identification systems. This causes different domains to resemble isolated islands with little to no connection, preventing the model from effectively integrating knowledge across domains to achieve better generalization.
Therefore, ID-based graph recommendation pre-training falls on two key issues: (1) how to well learn knowledge in each domain, and (2) how to effectively integrate all domains for joint training. To solve these issues, we design a hierarchical message-passing mechanism, which includes domain sub-graph propagation and cross-domain global graph propagation. 

\vspace{-0.2cm}  

\subsubsection{Domain Subgraph Propagation}
The fundamental task of pre-training is to effectively learn the knowledge of each domain. Since there are no user-item edges connecting different domains in recommender systems, each domain can actually be an independent subgraph. Thus, we first train models on each sub-graph. Following LightGCN and EDDA\cite{LightGCN:conf/sigir/0001DWLZ020,EDDA:conf/cikm/Ning0LCZT23}, we perform Grec\cite{EDDA:conf/cikm/Ning0LCZT23}, EDDA's core graph convolution, on source domains to capture fine-grained interaction patterns within the domain. Grec\cite{EDDA:conf/cikm/Ning0LCZT23} is formulated as:
\begin{small}
    \begin{equation}
\begin{aligned}
\mathbf{H}_s^{id(l+1)}& = (1-\alpha) \cdot \mathbf{\hat{A}} \mathbf{H}_s^{id(l)} + \alpha \cdot \mathbf{H}_s^{id(l)}  \\
\mathbf{\hat{A}} &= \mathbf{D}^{-1/2} \mathbf{A} \mathbf{D}^{-1/2},
\end{aligned}
\end{equation}
\end{small}
\vspace{-5pt}

where $\mathbf{\hat{A}} $ is the normalized adjacency matrix of each source domain sub-graph, and $\alpha=0.5$ balances the proportion of old and new representation.
    \vspace{-5pt}

\subsubsection{Cross-domain Global Graph Propagation}
After learning each subdomain's knowledge, the next step is to learn the global knowledge across all domains. However, as mentioned before, each domain is just like an isolated island with no edge connection; the information cannot diffuse across domains by edges.  A natural idea is that if some bridges can be established between these isolated islands, information propagation can be achieved.  Fortunately, in each domain, most items usually have textual information, which is universal. Users can also be described by their interacted items. We can use this textual information as a bridge to establish connections between each sub-domain. 

Recent work (e.g., AlphaRec\cite{AlphaRec25}) shows that semantic similarity in language model embeddings often aligns with user behavior patterns, suggesting a structural correspondence, though not equivalence, between language and behavior spaces in recommendation tasks. Building on this insight and later visualization case study (section \ref{sec_vis_case}), we explicitly leverage semantic similarity to facilitate ID embedding transfer across domains. Specifically, we construct a cross-domain semantic graph $\mathcal{G}_{global}^{pre}$ by generating semantic edges through text similarity, building a global graph connecting each pre-training domain.
We first process the textual information into a sentence. Then, considering the Large Language Model (LLM) has the world knowledge, we input the processed sentence into an LLM. It can be formulated as:
\begin{small}
    \begin{equation}
 x^{text}_v=LLM(c^{text}_v), x^{text}_u=LLM(c^{text}_u),
\end{equation}
\end{small}
where the $c^{text}_u$ and $c^{text}_v$ are the processed textual information of user $u$  and item $v$, respectively. $x^{text}_u$ and $x^{text}_v$ are the representation of corresponding user and item.
After that, we build edges between the multi-domains:
\begin{small}
\begin{equation}
\mathcal{G}_{global}^{pre} = 
\underbrace{\bigcup_{i=1}^N \mathcal{G}_s^{(i)}}_{\text{Src Subgraphs}} \cup 
\underbrace{\left\{
\begin{aligned}
&(u_i,u_j) \mid \cos(\mathbf{x}_{u_i}^{text}, \mathbf{x}_{u_j}^{text}) > \gamma \\
&(v_i,v_j) \mid \cos(\mathbf{x}_{v_i}^{text}, \mathbf{x}_{v_j}^{text}) > \gamma 
\end{aligned}
\right\}}_{\text{Cross-Domain Semantic Edges}},
\end{equation}
\end{small}
where \(\gamma\) is a manually selected hyper-parameter.

In the global graph, 
we use Grec for cross-domain propagation to capture cross-domain semantic associations:
\begin{small}
    \begin{equation}
\mathbf{H}_{global}^{id(l+1)} = \text{Grec}(\mathcal{G}_{global}^{pre},\; \mathbf{H}_{global}^{id(l)}).
\end{equation}
\end{small}

Note that, to preserve domain-specific knowledge, we utilize a new embedding table in the global graph to store cross-domain collaborative information denoted as $\mathbf{H}_{global}^{id(l+1)} $.

To fully leverage the universality of text features, we integrate ID embeddings and text representations so that they can complement each other when ID information is of suboptimal quality.

Fusion is performed as:
\begin{small}
\begin{equation}
\begin{aligned}
\mathbf{h}_s &= \text{L2-Norm}(\mathbf{h}_s^{id}) + \text{L2-Norm}(\text{Adapter}_s(\mathbf{x}^{text})) \\
\mathbf{h}_{global} &= \text{L2-Norm}(\mathbf{h}_{global}^{id}) + \text{L2-Norm}(\text{Adapter}_{global}(\mathbf{x}^{text})),
\end{aligned}
\end{equation}
\end{small}

where the Adapter module applies:
\begin{small}
\begin{equation}
\text{Adapter}(\mathbf{x}) = \mathbf{W}_{up} \cdot \text{ReLU}(\mathbf{W}_{down} \cdot \mathbf{x}),
\end{equation}
\end{small}
where $\mathbf{W}_{down}$ and $\mathbf{W}_{up}$ are learnable projection matrices for dimension reduction and restoration.
Through this hierarchical graph propagation, we obtain two types of embeddings: the Domain Subgraph Embedding $\mathbf{H}_s \in \mathbb{R}^{(|\mathcal{U}|+|\mathcal{V}|) \times d}$ and the Global Graph Embedding $\mathbf{H}_{global} \in \mathbb{R}^{(|\mathcal{U}|+|\mathcal{V}|) \times d}$. The domain subgraph embedding captures domain-specific interaction patterns, while the global graph embedding models cross-domain semantic associations. To construct a comprehensive embedding representation, we concatenate these two embeddings, resulting in 
\begin{small}
\begin{equation}
\mathbf{H}^{final} = \text{Concat}(\mathbf{H}_s, \mathbf{H}_{global}) \in \mathbb{R}^{(|\mathcal{U}|+|\mathcal{V}|) \times 2d}.
\end{equation}
\end{small}

This design provides several key benefits. Feature decoupling ensures that the domain subgraph embedding and the global graph embedding independently preserve local and global interaction patterns. We adopt simple addition for fusion as the hierarchical graph propagation itself effectively captures ID-text interactions, making the framework's core contribution robust regardless of fusion choice (see Appendix for comparisons with gated fusion and cross-attention). 
\vspace{-5pt}
\subsubsection{Pre-training Optimization} Following previous works\cite{LightGCN:conf/sigir/0001DWLZ020, EDDA:conf/cikm/Ning0LCZT23, 22NGCF:conf/sigir/Wang0WFC19}, the optimization objective during the pre-training phase includes BPR loss\cite{bpr:conf/uai/RendleFGS09} and regularization terms:
\begin{small}
    \begin{equation}
\mathcal{L}_{pre} = \sum_{i=1}^N \mathcal{L}_{BPR}^{(i)} + \lambda \left( \sum_{u} \|\mathbf{h}_u^{final}\|_2^2 + \sum_{v} \|\mathbf{h}_v^{final}\|_2^2 \right),
\end{equation}
\end{small}

where the BPR loss\cite{bpr:conf/uai/RendleFGS09} is defined as:
\begin{small}
    \begin{equation}
\mathcal{L}_{BPR} = \sum_{(u,v^+,v^-)} -\log\left(\sigma({\mathbf{h}_u}^{T} \mathbf{h}_{v^+} - {\mathbf{h}_u}^T \mathbf{h}_{v^-})\right).
\end{equation}
\end{small}
\subsection{Downstream Domain Transfer Fine-tuning Phase}
After the pre-training stage, we get a pre-trained graph recommendation model $f_g(G|\Theta)$. As previously mentioned, pre-trained graph recommendation focuses on transferring two key elements: the pre-trained parameters (GNN and ID embedding) and the graph structure.  However, 
 as the mismatching between the IDs in multi-domain pre-training and the IDs in downstream domains, the pre-trained model still cannot be directly applied to downstream recommendation. 
Intuitively, this problem can be resolved if we can establish a mapping relationship between the downstream domain and the pre-training domain. 
In recommender systems, similar items or users may exhibit similar behaviors, even across different domains. Therefore, similar to the pre-training stage, we can leverage textual information to map downstream users or items to pre-training domain users or items. For downstream data, if a user or item is similar to a pre-training user or item in textual semantic information, we create an edge between them.  In this way, we can both transfer the knowledge in pre-trained parameters and the structure information in pre-training graph.
\vspace{-6pt}


\subsubsection{Hierarchical Graph Knowledge Transfer}
In order to both transfer domain-specific knowledge and global multi-domain knowledge, we also design a hierarchical graph knowledge transfer mechanism, which includes local graph knowledge transfer and global knowledge transfer.

\paragraph{1. Local Graph Knowledge Transfer}
\textbf{Cross-domain Local Graph Construction}:
Explicitly establish semantic connections between the target domain and source domains:

\begin{small}
    \vspace{-10pt}
\begin{equation}
\mathcal{G}_{cross} = 
\underbrace{\mathcal{G}_t}_{\text{Tgt Subgraph}} \cup 
\underbrace{\bigcup_{i=1}^N \mathcal{G}_s^{(i)}}_{\text{Src Subgraphs}} \cup 
\underbrace{\left\{
\begin{aligned}
&(u_t,u_s) \mid \cos(\mathbf{x}_{u_t}^{text},\mathbf{x}_{u_s}^{text}) > \gamma \\
&(v_t,v_s) \mid \cos(\mathbf{x}_{v_t}^{text},\mathbf{x}_{v_s}^{text}) > \gamma 
\end{aligned}
\right\}}_{\text{Src-Tgt Semantic Edges}}.
\end{equation}
\end{small}

\textbf{Graph Propagation Initialization}:
Perform feature propagation in the cross-domain local graph:
\begin{small}
    \begin{equation}
\begin{bmatrix}
\mathbf{H}_t^{id} \\ 
\mathbf{H}_s^{id}
\end{bmatrix} = \text{Grec}\left(
\mathcal{G}_{cross},\;
\begin{bmatrix}
\mathbf{H}_t^{id} \\ 
\mathbf{H}_s^{id}
\end{bmatrix}
\right).
\end{equation}
\end{small}

\paragraph{2. Global Knowledge Transfer}
Construct an enhanced global graph to achieve global collaboration:
\begin{small}
    \begin{equation}
\mathcal{G}_{global}^{fine} = 
\underbrace{\mathcal{G}_{global}^{pre}}_{\text{Pre-train Graph}} \cup 
\underbrace{\mathcal{G}_t}_{\text{Tgt Subgraph}} \cup 
\underbrace{\left\{
\begin{aligned}
&(u_i,u_j) \mid \cos(\mathbf{x}_{u_i}^{text},\mathbf{x}_{u_j}^{text}) > \gamma \\
&(v_i,v_j) \mid \cos(\mathbf{x}_{v_i}^{text},\mathbf{x}_{v_j}^{text}) > \gamma 
\end{aligned}
\right\}}_{\text{Tgt Semantic Edges}}.
\end{equation}

\end{small}

Finally, we construct the enhanced global graph $\mathcal{G}_{global}^{fine}$ and perform global collaborative propagation:
\begin{small}
    \begin{equation}
\begin{bmatrix}
\mathbf{H}_{t,global}^{id} \\ 
\mathbf{H}_{s,global}^{id}
\end{bmatrix} 
= \text{Grec}\left(
\mathcal{G}_{global}^{fine},\;
\begin{bmatrix}
\mathbf{H}_{t,global}^{id} \\ 
\mathbf{H}_{s,global}^{id}
\end{bmatrix}
\right).
\end{equation}
\end{small}

Then, with hierarchical propagation, we fine-tune \modelname{} in the target domain to further adapt the learned representations and enhance recommendation performance.

\subsubsection{Fine-tuning Optimization}
In the fine-tuning stage, we also adopt BPR loss,
the optimization objective can be formulated as:
\begin{small}
\vspace{-6pt}
    \begin{equation}
\min_{\mathbf{H}_t^{id}, \Theta_{adapter_t}} \mathcal{L}_{BPR} + \eta \left( \sum_{u} \|\mathbf{h}_u^{final}\|_2^2 + \sum_{v} \|\mathbf{h}_v^{final}\|_2^2 \right),
\end{equation}
\end{small}

where the BPR loss \cite{bpr:conf/uai/RendleFGS09} is calculated based on the interaction data of the target domain.

\vspace{-6.5pt}

\begin{table*}[ht]
    \vspace{-6pt}
    \centering
    \small
    \caption{Results of different models in the cross-domain scenario.}
    \label{tab:cross_domain_models}
    \resizebox{0.91\textwidth}{!}{
    \begin{tabular}{p{3.5cm} l c c c c c c c c c c c c S }
    \toprule
    \textbf{Transfer Domains} & \textbf{Metric} & \textbf{DeepFM} & \textbf{AutoInt} & \textbf{LightGCN} & \textbf{DCN} & \textbf{PLE} & \textbf{STAR} & \textbf{MMOE} & \textbf{PEPNet} & \textbf{AlphaRec} & \textbf{EDDA} & \textbf{UniSRec} & \textbf{Ours} & \textbf{Rel Imp(\%)} \\
    \midrule
    \multirow{5}{*}{\parbox{3.5cm}{Automotive, Tools, Cell Phones, Clothing $\rightarrow$ Electronics}} 
    & AUC & 0.7278  & 0.7253 & 0.7220 & 0.7243 & 0.7291 & 0.7312 & 0.7288 & 0.7353 & 0.7227 & 0.7182 & \underline{0.7488} & \textbf{0.7789}$^*$ & \textbf{4.02\%} \\
    & Recall@10 & 0.3305 & 0.3302 & 0.3415 & 0.3328  & 0.3287 & 0.3292 & 0.3342 & 0.3347 & 0.3451 & 0.3583 & \underline{0.4002} & \textbf{0.4016} & \textbf{0.35\%} \\
    & Recall@20 & 0.4635 & 0.4581 & 0.4695 & 0.4664  & 0.4551 & 0.4515 & 0.4539 & 0.4562 & 0.4768 & 0.4397 & \underline{0.5227} & \textbf{0.5338}$^*$ & \textbf{2.12\%} \\
    & Precision@10 & 0.0577 & 0.0577 & 0.0597 & 0.0576 & 0.0592 & 0.0593 & 0.0597 & \underline{0.0602} & 0.0523 & 0.0557 & 0.0590 & \textbf{0.0641}$^*$ & \textbf{6.48\%} \\
    & Precision@20 & 0.0414 & 0.0410 & 0.0416 & 0.0414 & 0.0420 & 0.0416 & 0.0419 & \underline{0.0420} & 0.0369 & 0.0365 & 0.0393 & \textbf{0.0441}$^*$ & \textbf{5.00\%} \\
    \midrule
    
    \multirow{5}{*}{\parbox{3.5cm}{Automotive, Tools, Cell Phones, Clothing $\rightarrow$ Home}} 
    & AUC & 0.7178 & 0.7147 & 0.6953 & 0.7117 & 0.7052 & 0.7118 & 0.7073 & 0.7178 & 0.6946 & 0.6867 & \underline{0.7126} & \textbf{0.7352}$^*$ & \textbf{2.42\%} \\
    
    & Recall@10 & 0.3107 & 0.2988 & 0.3194 & 0.2990 & 0.2997 & 0.3062 & 0.3008 & 0.3051 & 0.2990 & 0.3176 & \textbf{0.3314} & \underline{0.3300} & \text{-} \\
    
    & Recall@20 & 0.4203 & 0.4290 & 0.4357 & 0.4233 & 0.4207 & 0.4279 & 0.4183 & 0.4325 & 0.4265 & 0.4396 & \underline{0.4747} & \textbf{0.4754} & \textbf{0.15\%} \\
    & Precision@10 & 0.0445 & 0.0413 & 0.0481 & 0.0456 & 0.0428 & 0.0442 & 0.0432 & 0.0447 & 0.0438 & 0.0504 & \underline{0.0514} & \textbf{0.0523}$^*$ & \textbf{1.75\%} \\
    & Precision@20 & 0.0290 & 0.0279 & 0.0303 & 0.0301 & 0.0284 & 0.0290 & 0.0281 & 0.0295 & 0.0318 & \underline{0.0344} & 0.0342 & \textbf{0.0371}$^*$ & \textbf{7.85\%} \\
    \midrule
    
    \multirow{5}{*}{\parbox{3.5cm}{Automotive, Tools, Cell Phones, Clothing $\rightarrow$ Movies}} 
    & AUC & 0.6972 & 0.6955 & 0.7008 & 0.6891 & 0.6985 & 0.7004 & 0.6984 & 0.6982 & 0.6889 & 0.6792 & \underline{0.7740} & \textbf{0.7912}$^*$ & \textbf{2.22\%} \\
    & Recall@10 & 0.2878 & 0.2821 & 0.3233 & 0.2836 & 0.3063 & 0.3041 & 0.3116 & 0.3041 & 0.3392 & 0.3567 & \textbf{0.4547} & \underline{0.4411} & \text{-} \\
    & Recall@20 & 0.3803 & 0.3963 & 0.4342 & 0.3998 & 0.4176 & 0.4139 & 0.4305 & 0.4127 & 0.4904 & 0.4847 & \underline{0.5413} & \textbf{0.5446} & \textbf{0.61\%} \\
    & Precision@10 & 0.0558 & 0.0557 & 0.0641 & 0.0561 & 0.0648 & 0.0647 & 0.0663 & 0.0644 & 0.0558 & 0.0757 & \underline{0.0798} & \textbf{0.0821} & \textbf{2.88\%} \\
    & Precision@20 & 0.0368 & 0.0384 & 0.0459 & 0.0384 & 0.0441 & 0.0442 & 0.0454 & 0.0437 & 0.0402 & \underline{0.0505} & 0.0486 & \textbf{0.0525}$^*$ & \textbf{3.96\%} \\
    \midrule
    
    \multirow{5}{*}{\parbox{3.5cm}{Automotive, Tools, Cell Phones, Clothing $\rightarrow$ Sports}} 
    & AUC & 0.6612 & 0.6659 & 0.6823 & 0.6741 & 0.6571 & 0.6646 & 0.6586 & 0.6761 & \underline{0.7061} & 0.6921 & 0.6922 & \textbf{0.7506}$^*$ & \textbf{6.30\%} \\
    & Recall@10 & 0.2422 & 0.2408 & 0.2650 & 0.2421 & 0.2203 & 0.2405 & 0.2277 & 0.2441 & 0.3132 & 0.3029 & \textbf{0.3647} & \underline{0.3618} & \text{-} \\
    & Recall@20 & 0.3546 & 0.3569 & 0.3816 & 0.3752 & 0.3347 & 0.3541 & 0.3425 & 0.3576 & 0.4568 & 0.4165 & \underline{0.4766} & \textbf{0.4981}$^*$ & \textbf{4.51\%} \\
    & Precision@10 & 0.0371 & 0.0394 & 0.0479 & 0.0376 & 0.0361 & 0.0396 & 0.0374 & 0.0401 & 0.0486 & \underline{0.0514} & 0.0501 & \textbf{0.0522}$^*$ & \textbf{1.56\%} \\
    & Precision@20 & 0.0284 & 0.0298 & 0.0343 & 0.0291 & 0.0277 & 0.0298 & 0.0285 & 0.0304 & 0.0339 & \underline{0.0360} & 0.0333 & \textbf{0.0365}$^*$ & \textbf{1.39\%} \\
\addlinespace[2pt]
\midrule
\midrule 
    \multirow{5}{*}{\parbox{3.5cm}{Books, Electronics $\rightarrow$ Clothing}} 
    & AUC & 0.6685 & 0.6658  & 0.6398 & 0.6650 & 0.6664 & 0.6659 & 0.6628 & 0.6676 & \underline{0.6873} & 0.6200 & 0.6474 & \textbf{0.6986}$^*$ & \textbf{1.64\%} \\
    & Recall@10  & 0.2458 & 0.2396 & 0.2204 & 0.2465 & 0.2379 & 0.2613 & \underline{0.2634} & 0.2445 & 0.2501 & 0.1994 & 0.1981 & \textbf{0.2681}$^*$ & \textbf{1.78\%} \\
    & Recall@20 & 0.3335 & 0.3305  & 0.3133 & 0.3379 & 0.3282 & \underline{0.3719} & 0.3719 & 0.3410 & 0.3513 & 0.2984 & 0.3239 & \textbf{0.3744} & \textbf{0.67\%} \\
    & Precision@10  & 0.0420 & 0.0428 & 0.0375 & \underline{0.0432} & 0.0422 & 0.0406 & 0.0410 & \underline{0.0432} & 0.0423 & 0.0384 & 0.0331 & \textbf{0.0489}$^*$ & \textbf{13.19\%} \\
    & Precision@20  & 0.0309 & 0.0302 & 0.0268 & 0.0306 & 0.0298 & 0.0296 & 0.0294 & 0.0307 & \underline{0.0309} & 0.0291 & 0.0274 & \textbf{0.0331}$^*$ & \textbf{7.12\%} \\
    \bottomrule
    \end{tabular}
    \vspace{-15pt}
    }
    \begin{flushleft}
        \scriptsize 
        \textbf{Note:} \textbf{Bold} and \underline{underlined} indicate the best and second-best performance, respectively. The symbol `$*$' denotes that the improvement over the best baseline is statistically significant with $p < 0.05$.
    \end{flushleft}
        \vspace{-15pt}

    \end{table*}

\section{Experiment}

In this section, we aim to answer the following research questions: How does \modelname{} perform compared to 
 other competitive baselines in cross-domain recommendation tasks (\textbf{RQ1})? Can the multi-domain pre-training of \modelname{} achieve competitive performance in multi-domain recommendation tasks (\textbf{RQ2})? How does \modelname{} perform in training-free scenarios where the model has not been trained on the target domain (\textbf{RQ3})? What are the effects of different model components on overall performance, as examined through an ablation study (\textbf{RQ4})? Does \modelname{} remain effective when applied to different model backbones (\textbf{RQ5})? Can \modelname{} maintain effectiveness under more challenging conditions, such as more lightweight LLMs, imperfect similarity graphs, and sparse or noisy textual inputs? (\textbf{RQ6})

\vspace{-5pt}

\subsection{Dataset}

In this study, we utilize the \textbf{Amazon Review Data (2018)} \cite{AmazonReview:conf/emnlp/NiLM19}, a widely-used benchmark for cross-domain recommendation research, known for its extensive user interaction records and rich semantic information, including product descriptions and user reviews.

To evaluate the model's adaptability across various domains and interaction densities, we organized the data into two collections (We provide more details in Appendix.\ref{app:datasets}).

\noindent (1) \textbf{8D (1 Year)}: This dataset spans one year up to August 15, 2018, covering interactions from eight domains: \textit{Automotive}, \textit{Tools and Home Improvement}, \textit{Cell Phones and Accessories}, \textit{Clothing, Shoes and Jewelry}, \textit{Electronics}, \textit{Home and Kitchen}, \textit{Movies and TV}, and \textit{Sports and Outdoors}. It includes data where each user or item has at least ten interactions, with a total of 1,148,521 interactions among 247,760 users and 107,245 items.

   \noindent(2) \textbf{3D (6 Months)}: This dataset focuses on three domains: \textit{Books}, \textit{Electronics}, and \textit{Clothing, Shoes and Jewelry} over six months, ending on August 15, 2018. It requires at least 20 interactions per user or item, with 524,876 interactions among 30,085 users and 30,851 items.
\vspace{-10pt}

\vspace{-5pt}

\paragraph{Data Processing}

Following established experimental methodologies from prior research \cite{UniCTR:journals/corr/abs-2312-10743, Aread:journals/corr/abs-2412-11905}, the datasets are processed to construct and extract key features. Categorical variables such as \texttt{userid}, \texttt{itemid}, \texttt{cateid}, \texttt{brand}, and \texttt{domain} are encoded, while continuous variables like \texttt{price} and \texttt{sales\_rank} are segmented into bins. Textual data from product \texttt{descriptions} and user \texttt{reviews} is also preprocessed to enhance the semantic profiles of the items.
\vspace{-10pt}

\subsection{Experimental Setting}

The dataset is divided into training (80\%), validation (10\%), and testing (10\%) sets based on timestamps, a common approach to simulate realistic application conditions \cite{Aread:journals/corr/abs-2412-11905}. Model performance is evaluated using three key metrics: AUC, Recall@K, and Precision@K, with \( K \in \{10, 20\} \).

Hyperparameter tuning is performed through grid search. The learning rate is selected from \( \{1 \times 10^{-3}, 5 \times 10^{-4}, 1 \times 10^{-4}, 5 \times 10^{-3}\} \), and the batch size is chosen from \( \{1024, 2048, 4096\} \), with the best configuration retained. $\gamma$ is selected from \( \{0.9, 0.95, 0.98, 0.99\} \). Following prior works\cite{EDDA:conf/cikm/Ning0LCZT23, DBLP:conf/pkdd/ZhengLJZLX24MICREC, DBLP:conf/cikm/NingCSHK0HLL022_RMSHRec} \textit{etc.}, the uniform sampling strategy is employed, where 100 negative items are uniformly sampled for each positive item in the test set. Models are trained using the Adam optimizer, with early stopping to prevent overfitting. 

\vspace{-8pt}
    

\subsection{Baselines}

In our experiments, we compare \modelname{} with several baselines. The single-domain models (DCN\cite{DCN:conf/kdd/WangFFW17}, DeepFM\cite{DeepFM:conf/ijcai/GuoTYLH17}, AutoInt\cite{Autoint:conf/cikm/SongS0DX0T19}, and LightGCN\cite{LightGCN:conf/sigir/0001DWLZ020}) are evaluated only within each target domain (for CDR tasks) or jointly on all domains (for MDR tasks), serving as reference methods without domain transfer.
Cross-domain models (MMOE\cite{MMOE:conf/kdd/MaZYCHC18}, PLE\cite{PLE:conf/recsys/TangLZG20}, PEPNet\cite{PEPNet:conf/kdd/ChangZHLNSG23}, STAR\cite{STAR:conf/cikm/ShengZZDDLYLZDZ21}, EDDA\cite{EDDA:conf/cikm/Ning0LCZT23}) are designed to transfer knowledge across domains and are tested in domain adaptation scenarios.
PLM-based recommenders such as UniSRec\cite{Unisrec:conf/kdd/HouMZLDW22} and AlphaRec\cite{AlphaRec25} use pre-trained language model embeddings to provide domain-invariant representations, further improving cross-domain generalization.

\subsection{Overall Performance Comparison (RQ1-RQ3)}

\begin{table*}[!t]
    \vspace{-4pt}
    \centering
    \small  
    \caption{Results in the multi-domain scenario.}
    \label{tab:pre-training_multi_domain}
    \resizebox{0.82\textwidth}{!}{
    \begin{tabular}{l c c c c c c c c c c c c S[table-format=2.2] }
    \toprule
    \textbf{Metric} & \textbf{DeepFM} & \textbf{AutoInt} & \textbf{LightGCN} & \textbf{DCN} & \textbf{MMOE} & \textbf{PLE} & \textbf{PEPNet} & \textbf{STAR} & \textbf{AlphaRec} & \textbf{EDDA} & \textbf{Ours} & \textbf{Rel Imp(\%)} \\
    \midrule
    \multicolumn{13}{c}{\textbf{8D (1 Year)}} \\
    \midrule
    
    AUC\textsubscript{Automotive} & 0.6389 & 0.6396 & 0.6323 & 0.6317 & 0.6356 & 0.6353 & 0.6394 & 0.6392 & \underline{0.6527} & 0.6458 & \textbf{0.6814}$^*$ & \textbf{4.40\%} \\
    
    AUC\textsubscript{Tools} & 0.7623 & 0.7639 & 0.7597 & 0.7581 & 0.7576 & 0.7619 & 0.7495 & \underline{0.7691} & 0.7359 & 0.7526 & \textbf{0.7846}$^*$ & \textbf{2.02\%} \\
    
    AUC\textsubscript{Cell Phones} & 0.7491 & \underline{0.7538} & 0.7145 & 0.7396 & 0.7386 & 0.7362 & 0.7489 & 0.7437 & 0.6784 & 0.7209 & \textbf{0.7553} & \textbf{0.20\%} \\

    AUC\textsubscript{Clothing} & 0.6927 & 0.6914 & 0.6712 & \underline{0.6951} & 0.6931 & 0.6937 & 0.6834 & 0.6744 &  0.6948 & 0.6851 & \textbf{0.7289}$^*$ & \textbf{4.86\%} \\

    AUC\textsubscript{Electronics} & 0.7281 & \underline{0.7334} & 0.7119 & 0.7300 & 0.7305 & 0.7296 & 0.7319 & 0.7312 & 0.7103 & 0.7278 & \textbf{0.7627}$^*$ & \textbf{4.00\%} \\

    AUC\textsubscript{Home} & 0.7107 & 0.7121 & 0.6953 & 0.7091 & 0.7046 & 0.7072 & \underline{0.7127} & 0.6946 & \underline{0.7127} & 0.6932 & \textbf{0.7214}$^*$ & \textbf{1.22\%} \\

    AUC\textsubscript{Movies} & 0.7055 & 0.6981 & 0.6413 & 0.7037 & 0.6983 & 0.6985 & 0.7093 & 0.7005 & 0.6792 & \underline{0.7331} & \textbf{0.7529}$^*$ & \textbf{2.70\%} \\

    AUC\textsubscript{Sports} & 0.6591 & 0.6662 & 0.6837 & 0.6655 & 0.6582 & 0.6576 & 0.6720 & 0.6621 & \underline{0.7061} & 0.6967 & \textbf{0.7410}$^*$ & \textbf{4.94\%} \\
    
    AUC\textsubscript{mean} & 0.6988 & \underline{0.7073} & 0.6888 & 0.6994 & 0.7020 & 0.7025 & 0.7059 & 0.7038 & 0.6955 & 0.7069 & \textbf{0.7410}$^*$ & \textbf{4.76\%} \\

    Recall@10\textsubscript{mean} & 0.2773 & 0.2800 & 0.3083 & 0.2766 & 0.2747 & 0.2758 & 0.2800 & 0.2795 & 0.3078 & \underline{0.3129} & \textbf{0.3276}$^*$ & \textbf{4.70\%} \\

    Recall@20\textsubscript{mean} & 0.3987 & 0.3988 & 0.4274 & 0.3992 & 0.3944 & 0.3952 & 0.4005 & 0.3984 & \underline{0.4356} & 0.4331 & \textbf{0.4620}$^*$ & \textbf{6.06\%} \\

    Precision@10\textsubscript{mean} & 0.0507 & 0.0511 & \underline{0.0541} & 0.0507 & 0.0506 & 0.0508 & 0.0507 & 0.0505 & 0.0460 & 0.0531 & \textbf{0.0548}$^*$ & \textbf{1.29\%} \\

    Precision@20\textsubscript{mean} & 0.0374 & 0.0373 & 0.0397 & 0.0374 & 0.0372 & 0.0373 & 0.0372 & 0.0370 & 0.0330 & \underline{0.0402} & \textbf{0.0404} & \textbf{0.50\%} \\
    
    \midrule
    \multicolumn{13}{c}{\textbf{3D (6 Months)}} \\
    \midrule
    
    AUC\textsubscript{Books} & 0.7818 & 0.8635 & 0.8936 & 0.8575 & 0.8605 & 0.7800 & 0.7796 & 0.8632 & 0.8927 & \underline{0.8940} & \textbf{0.9017} & \textbf{0.86\%} \\
    
    AUC\textsubscript{Electronics} & 0.5245 & 0.5288 & 0.5528 & 0.5680 & 0.6058 & 0.6006 & \underline{0.6191} & 0.5403 & 0.5776 & 0.5670 & \textbf{0.6826}$^*$ & \textbf{10.26\%} \\
    
    AUC\textsubscript{Clothing} & 0.6675 & 0.6611 & 0.6066 & 0.6732 & 0.6740 & 0.6664 & 0.6676 & \underline{0.6741} & 0.6198 & 0.6207 & \textbf{0.7007}$^*$ & \textbf{3.95\%} \\

    AUC\textsubscript{mean} & 0.6579 & 0.6845 & 0.6843 & 0.6996 & \underline{0.7134} & 0.6823 & 0.6888 & 0.6925 & 0.6966 & 0.6939 & \textbf{0.7617}$^*$ & \textbf{6.77\%} \\

    Recall@10\textsubscript{mean} & 0.2544 & 0.3811 & 0.4225 & 0.3818 & 0.3719 & 0.2549 & 0.2923 & 0.3894 & \underline{0.4498} & 0.4266 & \textbf{0.4546} & \textbf{1.07\%} \\

    Recall@20\textsubscript{mean} & 0.3911 & 0.5215 & 0.5511 & 0.5153 & 0.5079 & 0.3810 & 0.4307 & 0.5309 & \underline{0.5810} & 0.5583 & \textbf{0.5848} & \textbf{0.65\%} \\

    Precision@10\textsubscript{mean} & 0.0448 & 0.0715 & 0.0833 & 0.0711 & 0.0701 & 0.0439 & 0.0513 & 0.0728 & \underline{0.0890} & 0.0856 & \textbf{0.0898} & \textbf{0.90\%} \\

    Precision@20\textsubscript{mean} & 0.0355 & 0.0507 & 0.0571 & 0.0503 & 0.0503 & 0.0341 & 0.0398 & 0.0517 & \underline{0.0601} & 0.0586 & \textbf{0.0613}$^*$ & \textbf{2.00\%} \\
    \bottomrule
    \end{tabular}

    }
    \vspace{-1pt}
        \begin{flushleft}
        \scriptsize 
        \textbf{Note:} \textbf{Bold} and \underline{underlined} indicate the best and second-best performance, respectively. The symbol `$*$' denotes that the improvement over the best baseline is statistically significant with $p < 0.05$.
    \end{flushleft}
\end{table*}

\subsubsection{Cross-domain Scenario (RQ1)}

As shown in Table.\ref{tab:cross_domain_models}, our graph-based framework with ID transfer consistently improves cross-domain recommendation. \textbf{Models with Side Information and User Overlapping:}
Side information and user-overlapping models facilitate knowledge transfer but struggle in heterogeneous domains. For instance, PEPNet performs well on Books, Electronics $\rightarrow$ Clothing (AUC 0.6676) but underperforms on Automotive, Tools, Cell Phones, Clothing $\rightarrow$ Sports (AUC 0.6761).  \textbf{PLM-based Embedding Models:}
PLM-based models like UniSRec generally perform well in Recall@10, but often sacrifice overall Precision. While effective for semantic transfer, they lag in AUC and Precision compared to our approach. \textbf{Advantages of Our Method:}
Our similarity ID transfer mechanism better captures feature variations between source and target domains. On Automotive, Tools, Cell Phones, Clothing $\rightarrow$ Electronics, our model surpasses baselines by 4.02\% in AUC and 6.48\% in Precision@10.

\vspace{-8pt}

\subsubsection{Pre-training in Multi-domain Scenario (RQ2)}

As shown in Table.\ref{tab:pre-training_multi_domain}, pre-trained models improve key metrics such as AUC\textsubscript{mean} and Recall on both the \textbf{8D} and \textbf{3D} datasets. AUC\textsubscript{mean} increases by 4.76\% on \textbf{8D} and 6.77\% on \textbf{3D}.

\textbf{Impact of User Overlap and Data Sparsity:} User overlap and data sparsity are crucial factors for transfer learning. While pre-trained models benefit from high user overlap (such as 95\% in domains like \textit{Tools} and \textit{Cell Phones} in the \textbf{8D} dataset), our approach proves robust even in sparse scenarios. In the \textbf{3D} dataset, the \textit{Books} domain shares only 29.48\% of users with \textit{Electronics} and \textit{Clothing}, with no item overlap. Even under such challenging conditions, our models achieve significant improvements by leveraging graph-based similarity augmentation and semantic fusion (e.g., AUC\textsubscript{Electronics} improved by 10.26\%). These results demonstrate that our models effectively handle data scarcity and domain gaps.
\vspace{-5pt}

\subsubsection{Training-free Scenario (RQ3)}

\begin{figure*}[htbp]
    \centering    \includegraphics[width=0.87\textwidth]{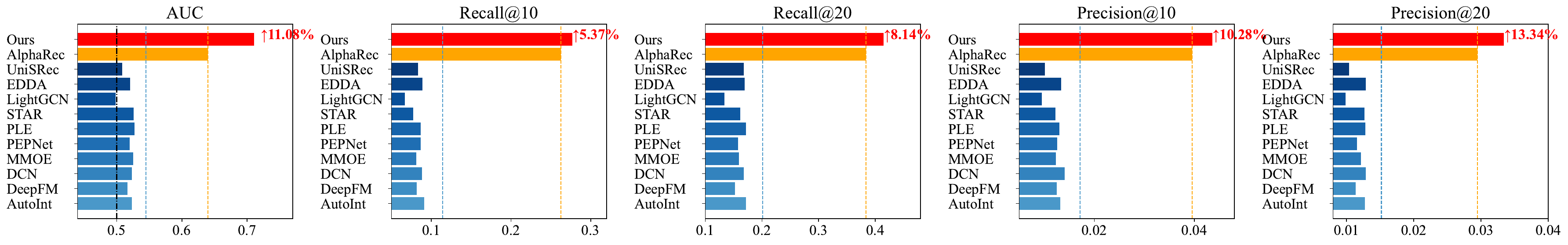}
    \caption{Training-free results on \textit{Automotive, Tools, Cell Phones, Clothing, Electronics, Home, Movies $\rightarrow$ Sports} dataset.}
    \label{fig:zeroshot_automotive_sports1}
\end{figure*}

We evaluate our method in a challenging training-free scenario by directly applying the pre-trained model to downstream data without fine-tuning. As shown in Fig.\ref{fig:zeroshot_automotive_sports1}, traditional transfer models like PEPNet are limited by sparse interactions and distributional shifts, while PLM-based methods such as AlphaRec and UniSRec benefit from semantic features but still fall short due to the lack of ID-level and hierarchical modeling. Our method consistently outperforms these baselines, demonstrating stronger generalization in zero-shot settings. Detailed results and additional experiments are presented in Appendix.\ref{app:zeroshot_results}.

\vspace{-4pt}

\subsection{Ablation Study (RQ4)}

\subsubsection{Cross-domain Ablation Study}

\begin{figure}[H]
    \vspace{-5pt}
    \centering
    \includegraphics[width=0.98\linewidth]{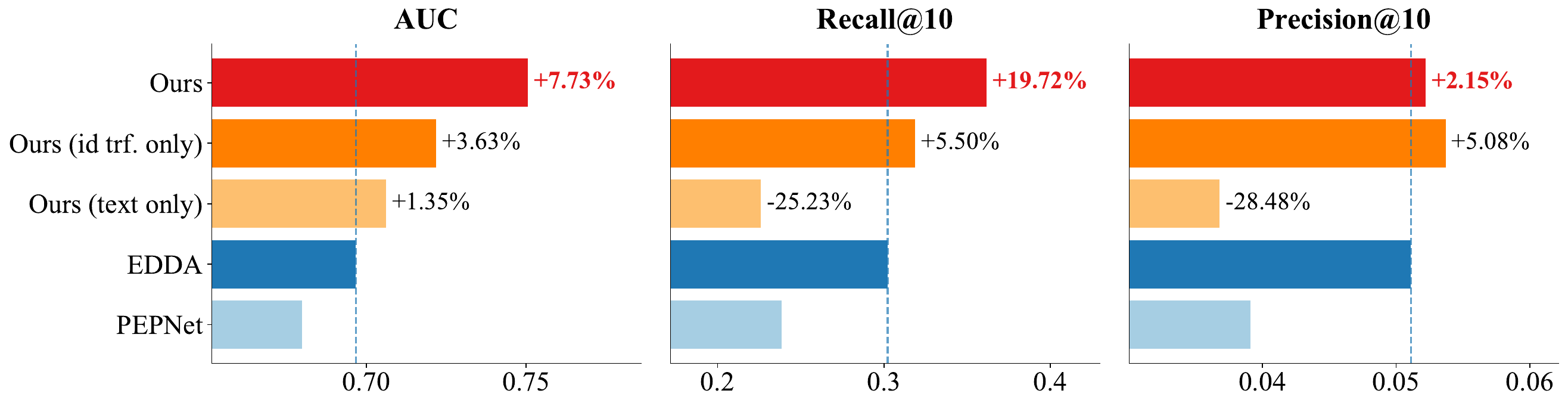}
    \caption{Cross-domain Ablation Study}
    \label{fig:performance_comparison_improvement}
    \vspace{-14pt}
\end{figure}

As shown in Figure.\ref{fig:performance_comparison_improvement}, in \textit{Automotive, Tools, Cell Phones, Clothing, Electronics, Home, Movies $\rightarrow$ Sports}, adding the ID transfer mechanism yields a 7.73\% AUC improvement over EDDA. A comparison of “Ours (id trf. only)” and “Ours (text only)” demonstrates that ID transfer features are essential: the former consistently outperforms the latter. Furthermore, our full model (integrating both ID transfer and semantic information) achieves notable improvements in AUC and Recall, with only a slight decrease in Precision. This highlights that while semantic information alone offers limited gains, its combination with ID transfer further enhances overall performance. We further analyze rare semantic-behavioral conflicts in Appendix.\ref{app:failure_case}.

\subsubsection{Multi-domain Ablation Study}

In 8D subset, as shown in Table.\ref{tab:ablation_results}, integrating graph structure augmentation, ID information, and textual information significantly enhances cross-domain transferability, yielding the best AUC and Recall@10 performance.

Removing graph semantic similar edge augmentation notably reduces Recall@10, highlighting its importance for capturing shared collaborative signals. Excluding ID information leads to substantial drops in both Recall and Precision, confirming its key role in modeling cross-domain relationships. In contrast, omitting textual information has minimal impact on Recall@10 and slightly improves Precision@10, indicating its effect is data-dependent: it may introduce noise, but remains useful in data-scarce scenarios. Under ID-only settings, our model surpasses PEPNet and EDDA in cross-domain transferability.
\vspace{-4pt}

\begin{table}[H]
\vspace{-8pt}
    \centering
    \caption{Multi-domain Ablation Study}
    \label{tab:ablation_results}
    \resizebox{0.9\linewidth}{!}{
        \begin{tabular}{lccc}
            \toprule
            Model & AUC\textsubscript{mean} & Recall@10 & Precision@10  \\
            \midrule
            Ours & 0.7410 & 0.3276 & 0.0548 \\
            w/o sim aug & 0.7385 (-0.34\%) & 0.3042 (-6.88\%) & 0.0552 (+0.93\%) \\
            w/o id & 0.6841 (-7.64\%) & 0.1829 (-44.13\%) & 0.0302 (-44.61\%) \\
            w/o text & 0.7273 (-1.80\%) & 0.3310 (+1.07\%) & 0.0612 (11.42\%) \\
            EDDA & 0.7069 (-4.56\%) & 0.3129 (-4.49\%) & 0.0531 (-3.23\%) \\
            \midrule
            PEPNet & 0.7059 (-4.69\%) & 0.2800 (-14.54\%) & 0.0507 (-7.40\%) \\

            \bottomrule
        \end{tabular}
    }
\vspace{-4pt}
\end{table}

\subsubsection{Cross-domain Training-free Ablation Study}

\vspace{-3pt}

As shown in Table.\ref{tab:performance_comparison_improvement_zero_shot}, EDDA (id only) backbone results remains modest. Ours (id trf. only) by only adding id transfer mechanism, outperforms EDDA and AlphaRec with an AUC of 0.6496, demonstrating the impact of ID transfer. Textual features alone (Ours text only) show moderate gains, consistent with AlphaRec's findings on semantic effectiveness.
The best performance is achieved by Ours (id trf. + text) with an AUC of 0.7106 (+36.44\%), highlighting the synergy between ID transfer and semantic fusion.


\begin{table}[h]
\vspace{-4pt}
    \centering
    \caption{Training-free Ablation Study}
    \label{tab:performance_comparison_improvement_zero_shot}
    \resizebox{0.9\linewidth}{!}{
        \begin{tabular}{lccc}
            \toprule
            Model & AUC & Recall@10 & Precision@10 \\
            \midrule
            EDDA & 0.5208 & 0.0890 & 0.0134 \\
            Ours (text only) & 0.6453 (+23.88\%) & 0.1502 (+68.80\%) & 0.0224 (+67.92\%) \\
            Ours (id trf. only) & 0.6496 (+24.74\%) & 0.2634 (+194.19\%) & 0.0434 (+224.70\%) \\
            Ours & 0.7106 (+36.44\%) & 0.2768 (+211.01\%) & 0.0436 (+225.37\%) \\
            \midrule
            AlphaRec & 0.6397 & 0.2628 & 0.0395 \\
            \bottomrule
        \end{tabular}
    }
        \vspace{-10pt} 
\end{table}

\subsection{Method Universality of \modelname{} (RQ5)}
\vspace{-2pt}

\begin{figure}[H]
    \vspace{-10pt}
        \centering
        \resizebox{0.8\linewidth}{!}{
            \begin{minipage}{\linewidth}
                \begin{subfigure}[t]{0.48\linewidth}
                    \centering
                    \includegraphics[width=\linewidth]{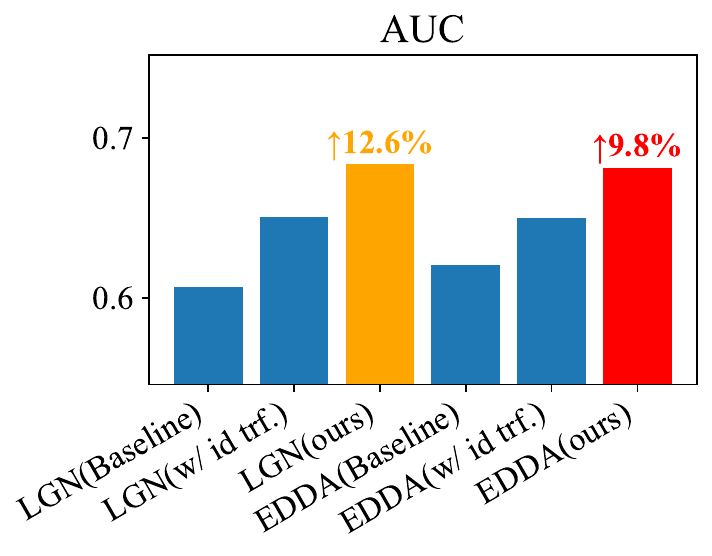}
                    \label{fig:subfig1_cross_auc}
                \end{subfigure}
                \hfill
                \begin{subfigure}[t]{0.48\linewidth}
                    \centering
                    \includegraphics[width=\linewidth]{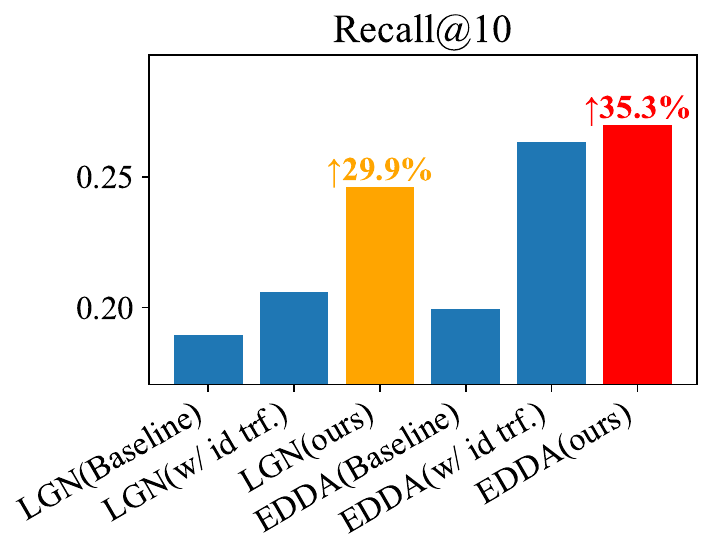}
                    \label{fig:subfig2_cross_recall}
                \end{subfigure}
            \end{minipage}
        }
        \vspace{-8pt}
        \caption{\small Universality: cross-domain scenario on 3D (↑\% shows relative improvement over Baseline)}
        \label{fig:overall_lgn}
        \vspace{-2pt}
    \end{figure}

We evaluate the universality of \modelname{} by applying it to LightGCN and EDDA. As shown in Figure.\ref{fig:overall_lgn}, ID transfer consistently improves performance across both base models, with AUC and Recall@10 increasing by up to 7.3\% and 19.5\% in cross-domain settings. Similar improvements are observed in training-free scenarios. These results demonstrate the effectiveness and strong generalization ability of \modelname{} across different models and scenarios. More comprehensive results are provided in Appendix.\ref{app:universality}.

\vspace{-6pt}
\subsection{Robustness and Sensitivity Analysis of \modelname{} (RQ6)}
This section evaluates the robustness of \modelname{} under challenging conditions, including lightweight LLMs, varying graph thresholds, noisy or missing text, and cold-start settings. \textbf{More detailed results and analysis are provided in Appendix.\ref{app:robustness}.}
\vspace{-4pt}
\subsubsection{Impact of LLM Capacity on Performance}

We evaluate the impact of language model size on representation quality and downstream performance. Comparing small models (e.g., BERT-110M) with larger ones (e.g., Llama-8B, SFR-Mistral-7B), we find that larger models offer slight improvements, while smaller models still perform well. Even BERT-110M achieves AUC=0.7273, surpassing all baselines with full information. This indicates that \modelname{} does not heavily depend on model size. The framework is agnostic to the embedding model; the LLM is used offline (one-time cost) and can be replaced with any encoder, including smaller models specifically trained for semantic similarity (e.g., top-ranked on MTEB), which are expected to perform even better than general-purpose LLMs at comparable sizes.

\begin{table}[ht]
\vspace{-6pt}
    \centering
    \caption{Impact of LLM Capacity on Performance}
    \resizebox{0.9\linewidth}{!}{
    \begin{tabular}{lcc}
        \toprule
        \textbf{Model} & \textbf{AUC} & \textbf{Recall@10} \\
        \midrule
        BERT-110M                & 0.7273 & 0.332 \\
        GPT2-medium-345M         & 0.7408 & 0.3418 \\
        Llama-8B                 & 0.7506 & 0.3618 \\
        SFR-Embedding-Mistral-7B & 0.7579 & 0.3644 \\
        \bottomrule
    \end{tabular}
    }
    \label{tab:llm_auc_main}
    \vspace{-8pt}
\end{table}
\subsubsection{Effect of Similarity Threshold \boldmath$\gamma$ on Structural Quality}

We analyze the impact of the similarity filtering threshold $\gamma$ on the quality of the constructed graph and downstream performance. Higher $\gamma$ values improve Recall and AUC by reducing noisy semantic edges; however, excessively high thresholds may degrade performance by filtering out useful semantic signals. A balanced $\gamma$ (e.g., 0.99) achieves the best overall results.
\begin{table}[htbp!]
\vspace{-4pt}
    \centering
    \caption{Impact of different $\gamma$ values on model performance.}
    \resizebox{0.9\linewidth}{!}{
    \begin{tabular}{lcccc}
        \toprule
        \textbf{Metric} & $\gamma = 0.9$ & $\gamma = 0.99$ & $\gamma = 0.995$ & $\gamma \in [0.6, 0.7]$ \\
        \midrule
        AUC       & 0.7490 & 0.7561 & 0.7511 & 0.7366 \\
        Rec@10    & 0.3249 & 0.3382 & 0.3359 & 0.3222 \\
        Prec@10   & 0.0548 & 0.0570 & 0.0561 & 0.0478 \\
        \bottomrule
    \end{tabular}
    }
    \label{tab:gamma_performance_main}
\end{table}

\vspace{-3pt}
\subsubsection{Robustness to Low-Quality or Noisy Textual Inputs}
\label{subsec:robustness}

We evaluate the robustness and sensitivity of \modelname{} to low-quality or missing textual information from two perspectives. First, our previous experiments on the real-world 8D Amazon dataset (which naturally includes noise and sparsity in key fields like \textit{features} (0.17\% missing), \textit{salesRank} (0.31\%), and \textit{brand} (18.75\%)) show that our model still achieves strong results (AUC 0.7561, Recall@10 0.3582), demonstrating resilience to sparse inputs. Second, controlled masking experiments (Table.\ref{tab:masking_simulation_main}) reveal that removing reviews leads to the largest drop, but our model still outperforms UniSRec (with full text). In contrast, masking IDs, titles, or numeric fields (e.g., price, salesRank) only slightly affects performance.
\begin{table}[ht]
    \vspace{-2pt}
    \centering
    \caption{Robustness under different types of masked information.}
    \resizebox{0.9\linewidth}{!}{
    \begin{tabular}{lcc}
        \toprule
        \textbf{Model / Mask Rate (\%)} & \textbf{AUC} & \textbf{Recall@10} \\
        \midrule
        Ours (full input)               & 0.7561 & 0.3582 \\
        - ID information (50\%)         & 0.7542 & 0.3548 \\
        - Reviews (50\%)                & 0.7261 & 0.3226 \\
        - Titles (50\%)                 & 0.7523 & 0.3558 \\
        - Descriptions, features (50\%) & 0.7515 & 0.3529 \\
        - Price, salesRank (50\%)       & 0.7507 & 0.3480 \\
        \midrule
        UniSRec (full info)             & 0.6924 & 0.3023 \\
        PEPNet (full info)              & 0.6967 & 0.3022 \\
        AlphaRec (full info)            & 0.7031 & 0.3132 \\
        \bottomrule
    \end{tabular}
    }
    \label{tab:masking_simulation_main}
\end{table}

\subsubsection{Adaptability to Cold-Start Domain-Adaptation Tasks}

To further evaluate generalizability, we conduct cold-start experiments on the target domain (Sports) by simulating a setting where only 5\% of target-domain interactions are available for training. As shown in Table.\ref{tab:coldstart_results_main}, most models suffer substantial performance degradation under such sparse supervision. In contrast, \modelname{} consistently achieves the best performance in both settings, significantly outperforming strong baselines such as AlphaRec and UniSRec. And even in the training-free setting, our model outperforms most fully trained baselines, highlighting its strong cross-domain transferability and robustness to severe data sparsity.

\begin{table}[htpb]
    \vspace{-4pt}
    \centering
    \caption{Cold-start results on the Sports domain}
    \resizebox{0.9\linewidth}{!}{
    \begin{tabular}{lccc}
    \toprule
    \textbf{Method}                    & \textbf{AUC} & \textbf{Recall@10} & \textbf{Precision@10} \\
    \midrule
    UniSRec (fully trained)           & 0.5328       & 0.0787             & 0.0103                \\
    EDDA (fully trained)    & 0.5215       & 0.1034             & 0.0151                \\
    LightGCN (fully trained)          & 0.5022       & 0.0820             & 0.0114                \\
    AlphaRec (fully trained) & \underline{0.5591} & \underline{0.1211} & \underline{0.0162}        \\
    Ours (fully trained)    & \textbf{0.5723}       & \textbf{0.1379}             & \textbf{0.0190}                \\
    \textbf{Rel Imp(\%)}     & \textbf{+2.36\%} & \textbf{+13.87\%} & \textbf{+17.28\%} \\
    \midrule
    AlphaRec (train-free)      & \underline{0.5220} & \underline{0.1178} & \underline{0.0162}        \\
    Ours (train-free)         & \textbf{0.5424}       & \textbf{0.1256}             & \textbf{0.0168}                \\
    \textbf{Rel Imp(\%)}     & \textbf{+3.91\%} & \textbf{+6.62\%}  & \textbf{+3.70\%} \\
    \bottomrule
    \end{tabular}
    }
    \label{tab:coldstart_results_main}
    \end{table}

\vspace{-4pt}

\subsection{Embedding Visualization \& Semantic Case Study} \label{sec_vis_case}

\begin{figure}[h]
\vspace{-6pt}
    \centering
    \includegraphics[width=0.8\linewidth]{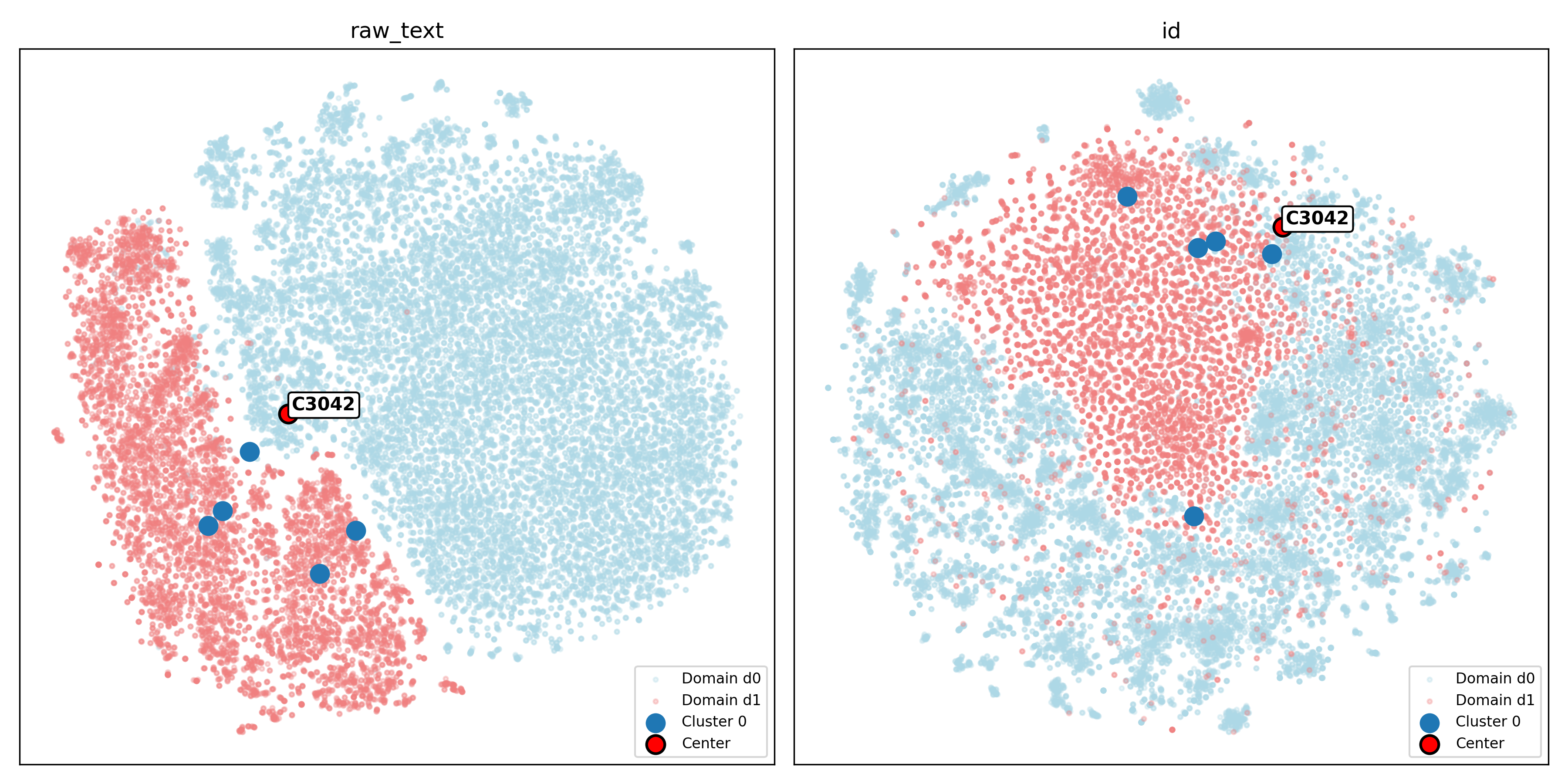}
    \caption{Embedding T-SNE in Books \& Clothes Domain}
    \label{fig1:tsn_book_mdr}
    \vspace{-14pt}
\end{figure}

To illustrate how semantic similarity aids ID embedding transfer, we analyze a case centered on the children’s book \textit{The Poky Little Puppy}. As shown in Figure.\ref{fig1:tsn_book_mdr}, text embeddings cluster this book with Clothes domain items (e.g., T-shirts, sweaters, hoodies, costumes, earrings) commonly linked to children. Item prompts indicate consistent gifting scenarios involving parents or grandparents. This semantic clustering also appears in the learned ID embedding space, where related cross-domain items remain close, suggesting that shared behaviors, like child gifting, are captured via semantic alignment. This finding supports and extends AlphaRec’s results\cite{AlphaRec25} from intra-domain to cross-domain. However, as recent studies note, semantic and collaborative features cannot fully replace each other, and how to best integrate both remains an open question\cite{MMGCN:conf/mm/WeiWHHC19, LLMRec:conf/wsdm/WeiRTWSCWYH24}.
\vspace{-3pt}

\section{Complexity Analysis}
Given $n$ domains (each with $N$ nodes and $E$ edges), the overall complexity includes $O(nNt_{\text{LLM}})$ for text embedding, $O(n^2NH)$ for similarity edge construction, and $O(nEH + n^2NH^2)$ for training. Considering $E$ and $N$ as primary variables, this simplifies to $O(E + N)$. Full details are in Appendix.\ref{app:complexity}.

\vspace{0.5em}
\noindent
\textbf{Efficiency.} On an RTX 3090 (24GB), \modelname{} uses about 12GB VRAM and completes each epoch in ~1 minute, while UniSRec requires about 20GB and ~6 minutes per epoch. The end-to-end cost is also competitive: overall training ($\to$Sports) takes ~34min for \modelname{} vs. ~32min for EDDA and ~15.3h for UniSRec. Semantic graph construction via Faiss-gpu takes ~40s for 107K items. New items only require embedding and incremental neighbor search without full graph rebuild (see Appendix for detailed breakdown). These results show \modelname{} maintains a favorable balance between effectiveness and efficiency.
\vspace{-4pt}

\section{Conclusion}

This paper presents \modelname{}, a graph neural network (GNN) framework for cross-domain recommendation. The method addresses key challenges in ID non-transferability and structural incompatibility across domains by using text as a semantic bridge while preserving collaborative signals. It integrates both graph-based and text-driven knowledge transfer through a pre-training and fine-tuning paradigm. Experiments across multi-domain, cross-domain, and training-free recommendation tasks demonstrate that \modelname{} achieves competitive performance, improving adaptability in settings with limited user overlap while avoiding the computational overhead of full PLM fine-tuning.

\section{Limitations and Future Work}

While \modelname{} demonstrates strong performance across multiple scenarios, two directions remain for future improvement. The similarity threshold $\gamma$ is currently set as a unified hyperparameter; exploring adaptive or learned thresholds across domain pairs could further optimize the semantic graph quality. The fusion of ID and text representations uses simple addition, which already achieves competitive results (see Appendix.\ref{app:fusion} for detailed comparisons); integrating more sophisticated fusion strategies into the GNN propagation layers may further enhance the model's capacity.

\bibliographystyle{ACM-Reference-Format}
\bibliography{main}

\newpage
\appendix
\renewcommand{\thefigure}{A.\arabic{figure}}
\setcounter{figure}{0}
\renewcommand{\thetable}{A.\arabic{table}}
\setcounter{table}{0}
\setcounter{page}{1}
\vspace{-5pt}

\clearpage

\section{Appendix}

\subsection{Notations}
\label{app:notations}

Table.\ref{tab:notations} summarizes the key notations used throughout this paper.

\begin{table}[H]
\centering
\small
\caption{Summary of key notations.}
\label{tab:notations}
\begin{tabular}{p{2.8cm} p{5.0cm}}
\toprule
\textbf{Notation} & \textbf{Description} \\ \midrule
\multicolumn{2}{l}{\textit{Graphs and Sets}} \\
$\mathcal{G}=(\mathcal{U},\mathcal{V},\mathcal{E})$ & Interaction graph; user set $\mathcal{U}$, item set $\mathcal{V}$, edges $\mathcal{E}$ \\
$\mathcal{G}_s^{(i)}$ & Interaction graph of the $i$-th source domain \\
$\mathcal{G}_t$ & Interaction graph of the target (downstream) domain \\
$\mathcal{G}_{global}^{pre}$ & Cross-domain global graph built in pre-training \\
$\mathcal{G}_{cross}$, $\mathcal{G}_{global}^{fine}$ & Cross-domain local / enhanced global graphs in fine-tuning \\
$N$ & Number of source domains \\ \midrule
\multicolumn{2}{l}{\textit{Embeddings}} \\
$\mathbf{X}_u^{text}$, $\mathbf{X}_v^{text}$ & Text feature matrices of users/items ($\in\mathbb{R}^{|\mathcal{U}|\times d_{text}}$, $\mathbb{R}^{|\mathcal{V}|\times d_{text}}$) \\
$\mathbf{x}_u^{text}$, $\mathbf{x}_v^{text}$ & Text representations of user/item from the LLM \\
$c_u^{text}$, $c_v^{text}$ & Processed textual information of user/item \\
$\mathbf{H}_s^{id}$, $\mathbf{H}_{global}^{id}$, $\mathbf{H}_t^{id}$ & ID embeddings on the source subgraph / global graph / target domain (layer index $l$ omitted) \\
$\mathbf{H}_s$, $\mathbf{H}_{global}$ & Fused subgraph / global embeddings after ID-text fusion ($\in\mathbb{R}^{(|\mathcal{U}|+|\mathcal{V}|)\times d}$) \\
$\mathbf{H}^{final}$ & Concatenated final embedding ($\in\mathbb{R}^{(|\mathcal{U}|+|\mathcal{V}|)\times 2d}$) \\
$\mathbf{h}_u^{final}$, $\mathbf{h}_v^{final}$ & Final per-user / per-item embeddings for recommendation \\ \midrule
\multicolumn{2}{l}{\textit{Operators and Functions}} \\
$\text{Grec}(\cdot)$ & Graph convolution operator (following EDDA) \\
$\text{Adapter}(\cdot)$ & Adapter module for text projection \\
$\hat{\mathbf{A}}$ & Symmetrically normalized adjacency ($\mathbf{D}^{-1/2}\mathbf{A}\mathbf{D}^{-1/2}$) \\
$\sigma(\cdot)$ & Sigmoid function \\
$f_g$, $f_t$ & Pre-trained GNN / target-domain recommendation function \\ \midrule
\multicolumn{2}{l}{\textit{Hyperparameters}} \\
$\gamma$ & Cosine-similarity threshold for semantic edges \\
$\alpha$ & Balance ratio between old and new representations ($=0.5$) \\
$d$, $d_{text}$ & Dimensions of ID / text embeddings \\
$\lambda$, $\eta$ & Regularization weights in pre-training / fine-tuning \\
\bottomrule
\end{tabular}
\end{table}

\subsection{Related Work} \label{app:related_work}

\subsubsection{Cross-domain Recommendation Methods}

\noindent\textbf{Single-domain Recommenders.}
Traditional models like DCN, DeepFM, and AutoInt\cite{DCN:conf/kdd/WangFFW17, DeepFM:conf/ijcai/GuoTYLH17, Autoint:conf/cikm/SongS0DX0T19} , rely on single-domain ID features and struggle with cross-domain transfer due to domain-dependent ID embeddings, even with auxiliary features like price or brand.

\noindent\textbf{Cross-domain Recommenders.}
Early methods such as PTUPCDR and EMCDR\cite{PTUPCDR:conf/wsdm/ZhuTLZXZLH22, EMCDR:conf/ijcai/ManSJC17} align overlapping IDs, but require strict user/item overlap. Others utilize kernel-based transfer and auxiliary signals like tags\cite{7cdr:journals/tnn/ZhangLWZ19, 8cdr:journals/tcyb/HaoZML19}, or adopt multi-task structures (e.g., MMOE, PLE, STAR\cite{MMOE:conf/kdd/MaZYCHC18, PLE:conf/recsys/TangLZG20, STAR:conf/cikm/ShengZZDDLYLZDZ21}) to share parameters across domains. Recent advances like PEPNet and subspace alignment\cite{PEPNet:conf/kdd/ChangZHLNSG23, 12cdr:conf/wise/ZhangLW018} improve transferability but remain constrained by explicit domain overlap. Pre-trained language model (PLM)-based methods, such as UniSRec and Uni-CTR\cite{Unisrec:conf/kdd/HouMZLDW22, UniCTR:journals/corr/abs-2312-10743}, pre-train on behavioral sequences and textual inputs. However, they either ignore user-item graph structures or require costly PLM fine-tuning, potentially impairing general semantics.

\subsubsection{Graph-based Cross-domain Recommendation}

\noindent\textbf{Single-domain Graph Models.}
GNN-based recommenders like NGCF, LightGCN, and Pinterest's large-scale model\cite{22NGCF:conf/sigir/Wang0WFC19, LightGCN:conf/sigir/0001DWLZ020, 26DBLP:conf/kdd/YingHCEHL18} improve collaborative filtering via message passing. SimGCL\cite{Xsimgcl:journals/tkde/YuXCCHY24} applies contrastive learning for representation robustness. Semantic-enhanced models like LATTICE and MMGCN\cite{LATTICE:conf/mm/Zhang00WWW21, MMGCN:conf/mm/WeiWHHC19} incorporate multimodal features, but struggle with ID transfer across domains. LLMRec\cite{LLMRec:conf/wsdm/WeiRTWSCWYH24} explores LLMs for graph augmentation, yet lacks mechanism for structural transfer.

\noindent\textbf{Cross-domain Graph Models.}
Existing cross-domain graph models leverage techniques like multi-task decoupling \cite{EDDA:conf/cikm/Ning0LCZT23}, metapath-guided aggregation \cite{M2GNN:conf/sigir/HuaiYZZL023}, inter-graph modeling \cite{14Cross-GraphKnowledgeTransferNetwork:conf/icc/00030W021}, attention-based transfer \cite{13DBLP:conf/cikm/LiuLLP20, 32:conf/ijcai/ZhuWCLZ20}, and reinforcement learning \cite{RLCDR:journals/tkde/LiHL24}. However, they face limitations; for instance, hypergraph models like II-HGCN \cite{IIGCN:conf/www/HanZCCY23} falter in low-overlap scenarios. Graph pre-training for cross-domain recommendation is underexplored. Early attempts either rely on structural overlap and weak MF fine-tuning \cite{20pretrainGraph_wang2021pre}. General pre-trained GNNs\cite{19:conf/kdd/QiuCDZYDWT20, GPPT:conf/kdd/SunZHWW22, gpt-gnn:conf/kdd/HuDWCS20} are not specifically optimized for cross-domain recommendation, leaving challenges like non-transferable ID embeddings and domain-specific graph structures largely unaddressed. More recently, AlphaRec \cite{AlphaRec25} showed generalization with text-only features but sidestepped the core challenge of transferring ID-based collaborative signals, leaving open the question of how to preserve and adapt them across domains.

\subsection{Datasets} \label{app:datasets}

In this study, we utilize the \textbf{Amazon Review Data (2018)} \cite{AmazonReview:conf/emnlp/NiLM19}, known for its extensive user interaction records and rich semantic information, including product descriptions and user reviews.

To evaluate the model's adaptability across various domains and interaction densities, we organized the data into two collections (We provide more details in Appendix.\ref{app:datasets}).

\noindent (1) \textbf{8D (1 Year)}: This dataset spans one year up to August 15, 2018, covering interactions from eight domains: \textit{Automotive}, \textit{Tools and Home Improvement}, \textit{Cell Phones and Accessories}, \textit{Clothing, Shoes and Jewelry}, \textit{Electronics}, \textit{Home and Kitchen}, \textit{Movies and TV}, and \textit{Sports and Outdoors}. It includes data where each user or item has at least ten interactions, with a total of 1,148,521 interactions among 247,760 users and 107,245 items.

   \noindent(2) \textbf{3D (6 Months)}: This dataset focuses on three domains: \textit{Books}, \textit{Electronics}, and \textit{Clothing, Shoes and Jewelry} over six months, ending on August 15, 2018. It requires at least 20 interactions per user or item, with 524,876 interactions among 30,085 users and 30,851 items.
\vspace{-10pt}
See Table.\ref{tab:datasets}.
\begin{table*}[h]
\centering
\caption{Statistics of Datasets from Two Collections}
\label{tab:datasets}
\begin{tabular}{lcccccc}
\hline
\textbf{Collection} & \textbf{Domain} & \textbf{Users} & \textbf{Items} & \textbf{Interactions} & \textbf{Density ($\times 10^{-3}$)} \\ \hline
\multirow{8}{*}{8D (1 Year)} 
& Automotive               & 20,860  & 9,896  & 85,713   & 0.415 \\
& Tools and Home Improvement & 16,486  & 4,676  & 52,227   & 0.677 \\
& Cell Phones and Accessories & 6,962   & 3,014  & 18,520   & 0.883 \\
& Clothing, Shoes and Jewelry & 58,982  & 30,160 & 331,866  & 0.187 \\
& Electronics              & 41,448  & 14,706 & 182,311  & 0.299 \\
& Home and Kitchen         & 61,303  & 25,466 & 293,047  & 0.188 \\
& Movies and TV            & 10,863  & 5,371  & 64,624   & 1.108 \\
& Sports and Outdoors      & 30,856  & 13,956 & 120,213  & 0.279 \\ \cline{2-6}
& \textbf{Summary}         & 247,760 & 107,245 & 1,148,521 & 0.043 \\ \hline
\multirow{3}{*}{3D (6 Months)} 
& Books                    & 18,566  & 18,417 & 449,458  & 1.314 \\
& Electronics               & 6,792   & 8,063  & 54,978   & 1.005 \\
& Clothing, Shoes and Jewelry & 4,727   & 4,371  & 20,440   & 0.990 \\ \cline{2-6}
& \textbf{Summary}         & 30,085  & 30,851  & 524,876  & 0.566 \\ \hline
\end{tabular}
\end{table*}

\subsection{Prompt-based Embedding Generation Details}
\label{appendix:prompt_details}

\begin{figure}[h!]
    \centering
    \includegraphics[width=0.74\linewidth]{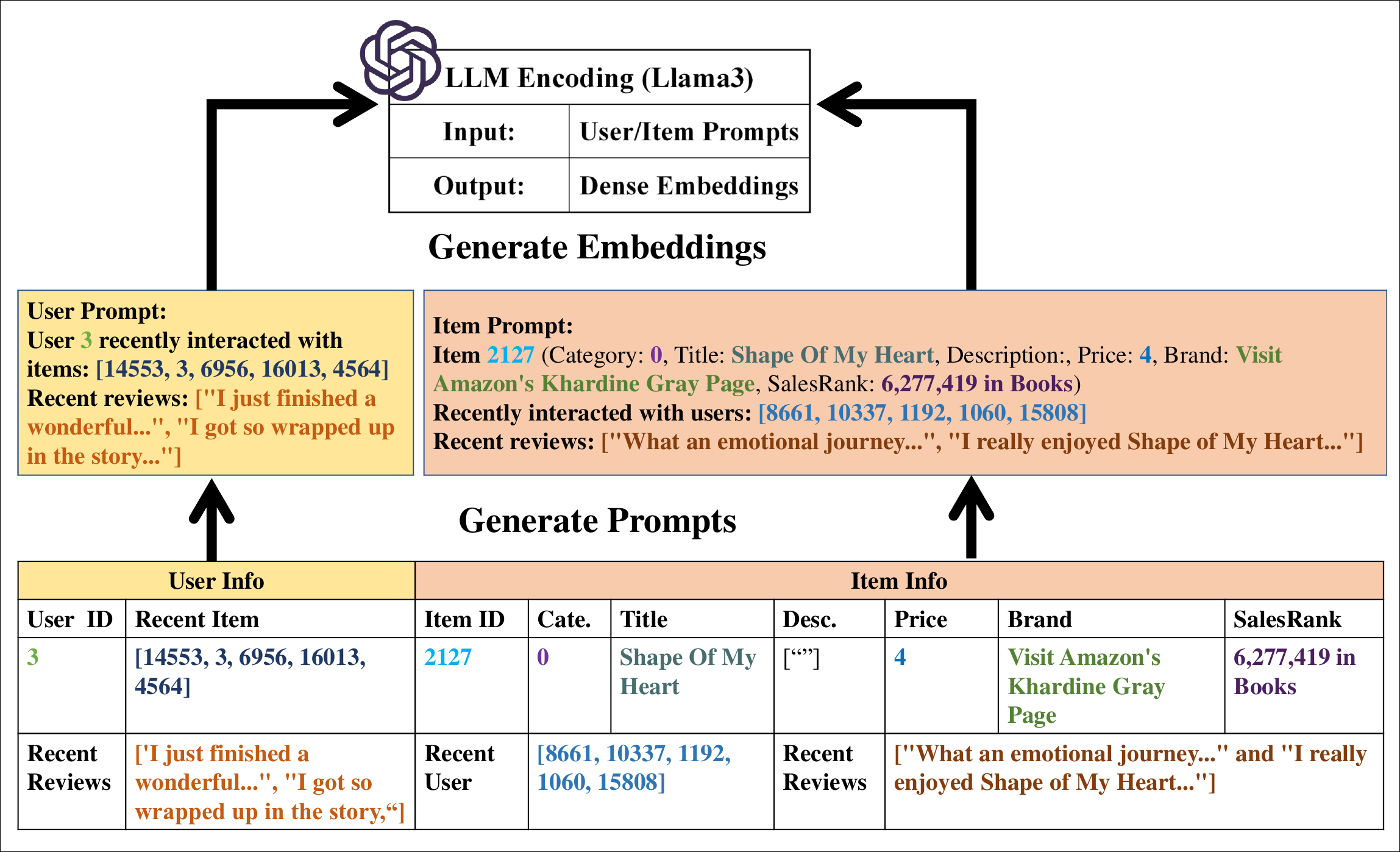}
    \caption{Illustration of the prompt generation pipeline.}
    \label{fig:prompt_pipeline}
\end{figure}

As shown in Figure.\ref{fig:prompt_pipeline}, we generate semantic embeddings $\mathbf{X}_u^{text} \in \mathbb{R}^{|\mathcal{U}| \times d_{text}}$ and $\mathbf{X}_v^{text} \in \mathbb{R}^{|\mathcal{V}| \times d_{text}}$ by leveraging large language models (LLMs) to encode textual and interaction-based information into dense vector representations. The pipeline consists of preprocessing textual data, summarizing interaction histories, generating prompts, and finally encoding them into semantic embeddings using models such as \texttt{Llama 3}~\cite{llama3modelcard}.
\begin{itemize}
    \item \textbf{Textual Data Preprocessing:} Rich textual information from product \texttt{descriptions} and user \texttt{reviews} was tokenized, truncated to a predefined maximum length, and encoded to build enhanced semantic profiles.
    \item \textbf{Interaction Histories:} Recent interactions were aggregated for both users and items: For each \texttt{userid}, the most recent $k$ items and their corresponding \texttt{reviews} were appended to create a historical interaction sequence from the training set. Similarly, for each \texttt{itemid}, the $k$ most recent interacting \texttt{users} and associated \texttt{reviews} were retrieved.
    \item \textbf{Prompt Generation and Truncation:} Prompts were formulated to summarize interaction histories and item attributes, providing an input suitable for LLM embedding generation. A quantile-based truncation approach was applied to cap prompt lengths and ensure computational efficiency.
    \item \textbf{LLM Embedding Generation:} Using models such as \texttt{Llama 3}~\cite{llama3modelcard}, the prompts were encoded into dense vector representations. These embeddings form the backbone of the semantic feature set.
\end{itemize}

\subsection{Baselines} \label{app:baselines}
In this section, we introduce the baseline models used for comparison in our experiments. In our experiments, the single-domain models (DCN, DeepFM, AutoInt, and LightGCN) provide a baseline performance that does not involve domain transfer. These models are used to assess the performance in a single, fixed domain. On the other hand, the cross-domain models (MMOE, PLE, PEPNet, STAR, EDDA) are designed to handle domain shifts, enabling them to generalize across different domains. These models are tested in scenarios where the goal is to transfer knowledge to the target domain, showcasing their capability to adapt and perform well across multiple domains. PLM-based cross-domain recommenders like UniSRec\cite{Unisrec:conf/kdd/HouMZLDW22} and AlphaRec\cite{AlphaRec25}, which leverage pre-trained language model(PLM) embeddings, further enhance cross-domain generalization by providing domain-invariant representations that facilitate knowledge transfer. The details of the mentioned baselines  are listed as follows:

\begin{itemize}
    \item \textbf{DCN} \cite{DCN:conf/kdd/WangFFW17} (KDD 2017):  
    DCN (Deep Cross Network) captures high-order feature interactions using cross-network layers. It is primarily designed for single-domain recommendation tasks. In our experiments, it serves as a baseline to assess performance in the target domain without domain transfer.

    \item \textbf{DeepFM} \cite{DeepFM:conf/ijcai/GuoTYLH17} (IJCAI 2017):  
    DeepFM (Deep Factorization Machine) combines factorization machines for capturing feature interactions with deep learning for nonlinear transformations. As a single-domain model, it is trained and tested on the target domain to evaluate its performance without domain transfer.

    \item \textbf{MMOE} \cite{MMOE:conf/kdd/MaZYCHC18} (KDD 2018):  
    MMOE (Multi-gate Mixture-of-Experts) is a multi-task learning model that uses multiple gate-controlled mixture-of-experts modules to manage shared and specific task information. It is selected for its ability to generalize across domains, optimizing multiple tasks simultaneously. This makes it suitable for transferring knowledge across domains, particularly in cross-domain recommendation tasks where domain-specific features need to be shared or adapted.

    \item \textbf{AutoInt} \cite{Autoint:conf/cikm/SongS0DX0T19} (CIKM 2019):  
    AutoInt (Automatic Feature Interaction Learning) uses self-attention mechanisms to automatically learn feature interactions. It is typically applied in single-domain scenarios, and here, it is tested directly on the target domain to provide a baseline performance.

    \item \textbf{PLE} \cite{PLE:conf/recsys/TangLZG20} (RecSys 2020):  
    PLE (Progressive Layered Extraction network) enhances feature representations through a progressive, step-by-step approach using expert and gate networks. It is chosen for its ability to balance shared and task-specific information, which is critical when transferring knowledge between domains. This makes it effective in cross-domain recommendation tasks, where models need to adapt to both shared and distinct domain features.

    \item \textbf{LightGCN} \cite{LightGCN:conf/sigir/0001DWLZ020} (SIGIR 2020):  
    LightGCN (Light Graph Convolutional Network) simplifies graph convolutional networks by focusing on collaborative filtering signals in user-item interaction graphs. It is primarily used for single-domain recommendation tasks, and here, we assess its performance in the target domain.

    \item \textbf{STAR} \cite{STAR:conf/cikm/ShengZZDDLYLZDZ21} (CIKM 2021):  
    STAR (Star Topology Adaptive Recommender) leverages a star topology to handle tasks across multiple domains simultaneously. By adapting its parameters to the characteristics of each domain, STAR captures both common and domain-specific information. This makes it suitable for cross-domain recommendation, as it is designed to handle domain shifts and learn domain-invariant representations.

    \item \textbf{UniSRec} \cite{Unisrec:conf/kdd/HouMZLDW22} (KDD 2022):  
    UniSRec (Universal Sequence Representation Learning for Recommender Systems) utilizes self-supervised learning to generate universal sequence representations that can be applied across different domains. Its ability to learn domain-invariant features makes it an ideal model for cross-domain tasks, where the goal is to transfer knowledge across multiple domains.

    \item \textbf{PEPNet} \cite{PEPNet:conf/kdd/ChangZHLNSG23} (KDD 2023):  
    PEPNet (Parameter and Embedding Personalized Network) dynamically adjusts embeddings and DNN parameters using personalized prior information. It is selected for its ability to handle variations in tasks and domains, effectively enabling domain adaptation and transfer by modifying embeddings and network structures according to domain-specific needs.

    \item \textbf{EDDA} \cite{EDDA:conf/cikm/Ning0LCZT23} (CIKM 2023):  
    EDDA (Embedding Disentangling and Domain Alignment) disentangles embeddings into generalizable and domain-specific components, enabling cross-domain knowledge transfer. It is particularly effective at handling domain shifts by aligning domain-specific features while retaining commonalities across domains, making it a strong candidate for cross-domain recommendation tasks.

    \item \textbf{AlphaRec} \cite{AlphaRec25} (ICLR 2025):  
    AlphaRec constructs graph recommendation models directly from item textual metadata without using ID embeddings. It employs pre-trained language model representations as input features, which are transformed via a lightweight architecture consisting of a multilayer perceptron, graph convolution, and contrastive learning. By eliminating reliance on ID information, AlphaRec learns transferable item representations and demonstrates strong generalization across domains. This makes it a representative text-driven graph-based method for domain generalized recommendation.
\end{itemize}

\subsection{Additional Training-free Results (RQ3)}
\label{app:zeroshot_results}

\begin{figure*}[ht!]
    \centering
    \includegraphics[width=0.9\textwidth]{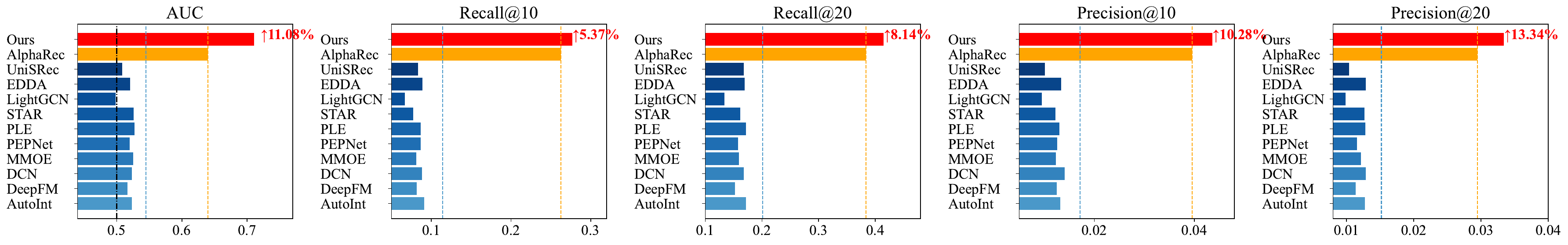}
    \caption{Training-free results on \textit{Automotive, Tools, Cell Phones, Clothing, Electronics, Home, Movies $\rightarrow$ Sports} dataset.}
    \label{fig:zeroshot_automotive_sports}
\end{figure*}

\begin{figure*}[htbp]
    \centering
    \includegraphics[width=0.9\textwidth]{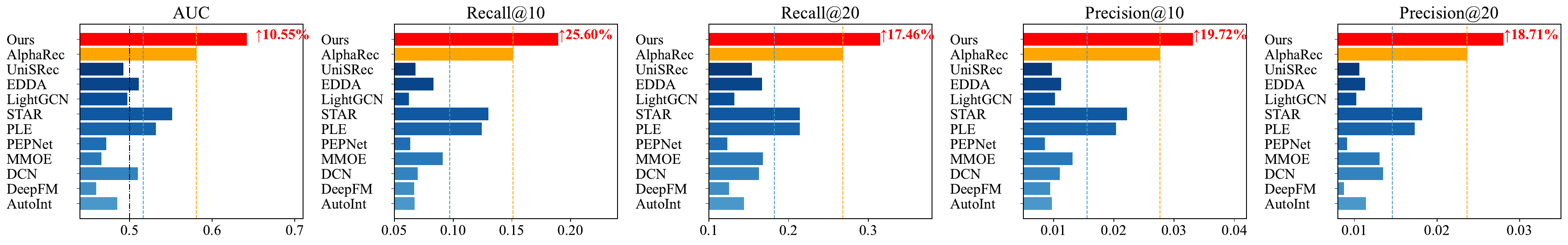}
    \caption{Training-free results on \textit{Books, Electronics $\rightarrow$ Clothing} dataset.}
    \label{fig:zeroshot_books_clothing}
\end{figure*}

In this section, we evaluate our method in a more challenging training-free setting. In this setting, we directly apply our pre-trained model to the downstream data without any fine-tuning. The results are shown in Fig.\ref{fig:zeroshot_books_clothing} and Fig.\ref{fig:zeroshot_automotive_sports}. From those figures, we have the following observations:

ID and side information based transfer models, such as PEPNet, rely on shared and specific features for cross-domain knowledge transfer but struggle with sparse interactions and distributional differences. On the \textit{Automotive, Tools, Cell Phones, Clothing, Electronics, Home, Movies $\rightarrow$ Sports} dataset (Figure.\ref{fig:zeroshot_automotive_sports}), PEPNet achieves an AUC of 0.5200. 

PLM-based models such as UniSRec and AlphaRec benefit from semantic representations learned from item text. AlphaRec performs much better than UniSRec by incorporating graph-based modeling and semantic information, achieving strong results in both AUC and Recall. However, due to the absence of ID-level behavior transfer and lack of hierarchical domain modeling, it still falls short of our method. For example, on the \textit{Books, Electronics $\rightarrow$ Clothing} dataset (Figure.\ref{fig:zeroshot_books_clothing}), AlphaRec achieves an AUC of 0.5808 and Recall@10 of 0.1509, while our method reaches 0.6420 and 0.1895, outperforming all baselines.

\subsection{Method Universality of \modelname{} (RQ5)}
\label{app:universality}

\begin{figure}[htbp]
\vspace{-10pt}
    \centering
    \resizebox{0.9\linewidth}{!}{ 
        \begin{minipage}{\linewidth}
            \begin{subfigure}[t]{0.48\linewidth}
                \centering
                \includegraphics[width=\linewidth]{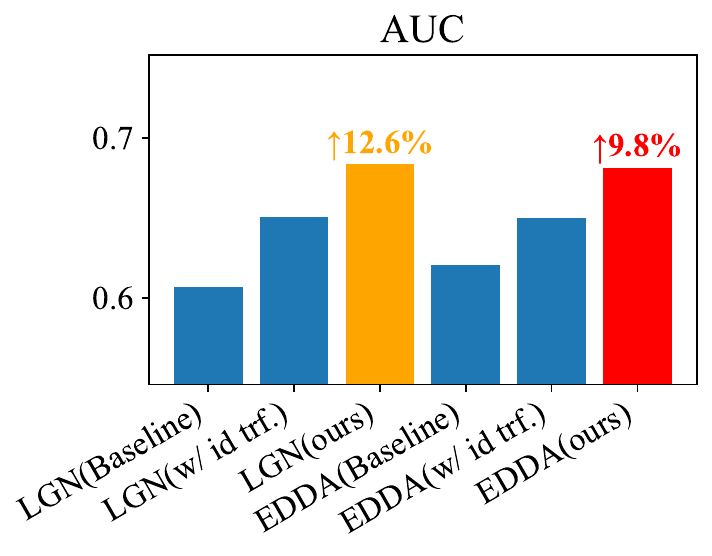}
                \label{fig:subfig1_cross_auc_app}
            \end{subfigure}
            \hfill
            \begin{subfigure}[t]{0.48\linewidth}
                \centering
                \includegraphics[width=\linewidth]{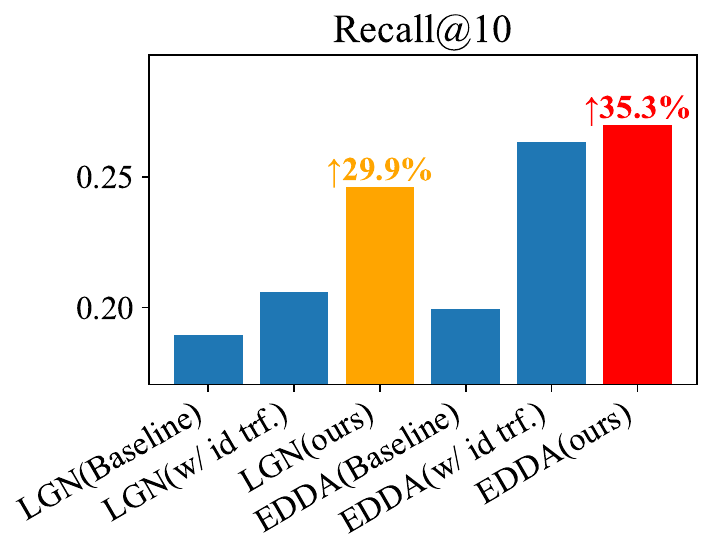}
                \label{fig:subfig2_cross_recall_app}
            \end{subfigure}
        \end{minipage}
    }
    \vspace{-6pt}
    \caption{\small Universality: cross-domain scenario on 3D (↑\% shows relative improvement over Baseline)}
    \label{fig:overall_lgn_app}
    \vspace{-2pt}
\end{figure}

\begin{figure}[htbp]
    \centering
    \vspace{-10pt}
    \resizebox{0.9\linewidth}{!}{ 
        \begin{minipage}{\linewidth}
            \begin{subfigure}[t]{0.48\linewidth}
                \centering
                \includegraphics[width=\linewidth]{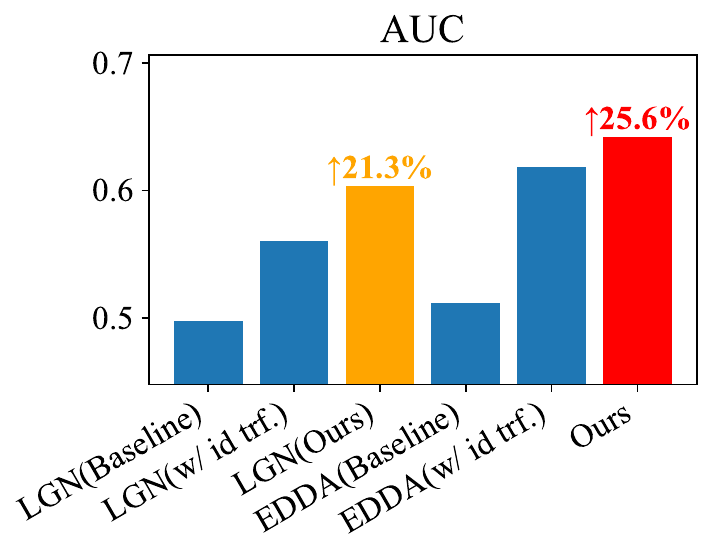}
                \label{fig:subfig1_zeroshot_auc}
            \end{subfigure}
            \hfill
            \begin{subfigure}[t]{0.48\linewidth}
                \centering
                \includegraphics[width=\linewidth]{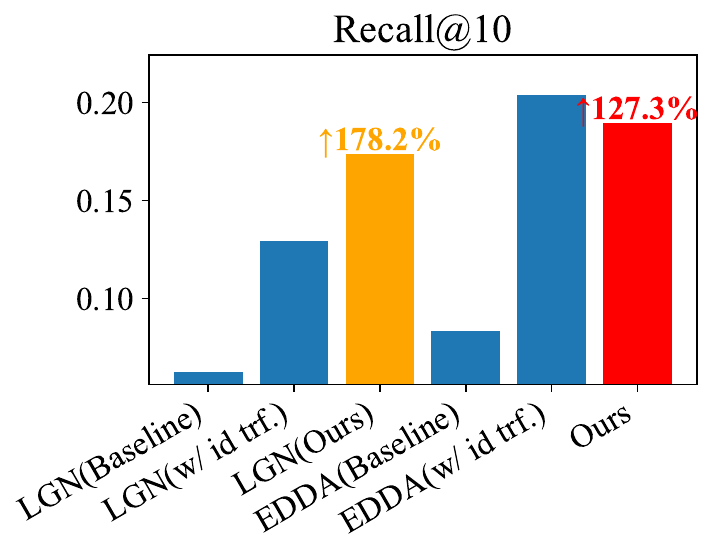}
                \label{fig:subfig2_zeroshot_recall}
            \end{subfigure}
        \end{minipage}
    }
    \vspace{-4pt}
    \caption{\small Universality: training-free scenario on 3D (↑\% shows relative improvement over Baseline)}
    \label{fig:overall_zeroshot}
\end{figure}

This section aims to evaluate the generality and effectiveness of the proposed \modelname{}. To achieve this, we apply our method to base models (LightGCN and EDDA), and Figures.\ref{fig:overall_lgn_app} and \ref{fig:overall_zeroshot} summarize the performance in cross-domain and training-free recommendation tasks, respectively.

In cross-domain recommendation (Figure.\ref{fig:overall_lgn_app}), ID transfer boosts AUC by 7.3\% for LightGCN and 4.7\% for EDDA, with further gains of 5.0\% and 4.9\% from adding textual features. Recall@10 shows similar improvements, rising by 8.7\% and 6.5\% with ID transfer, and by another 19.5\% and 2.5\% after incorporating text.

\vspace{-2pt}

In training-free recommendation (Figure.\ref{fig:overall_zeroshot}), ID transfer significantly improves AUC by 12.7\% on LightGCN and 20.8\% on EDDA, with additional gains from textual features (7.6\% and 3.9\%). Recall@10 sees even larger increases (107.3\% and 144.4\% from ID transfer), and an extra 34.3\% from text on LightGCN. Although textual features may slightly affect accuracy in zero-shot settings (as noted in the ablation study), overall performance remains strong, with a 127.3\% Recall gain over the baseline.

Together with cross-domain results, these findings confirm the generality of \modelname{}, where ID transfer drives major improvements and textual features enhance generalization across base models.

\subsection{Visualization \& Case Study}

\label{vis_case}
\subsubsection{Semantic Consistency within Book Domain}
\begin{figure}[h]
    \centering
    \includegraphics[width=0.95\linewidth]{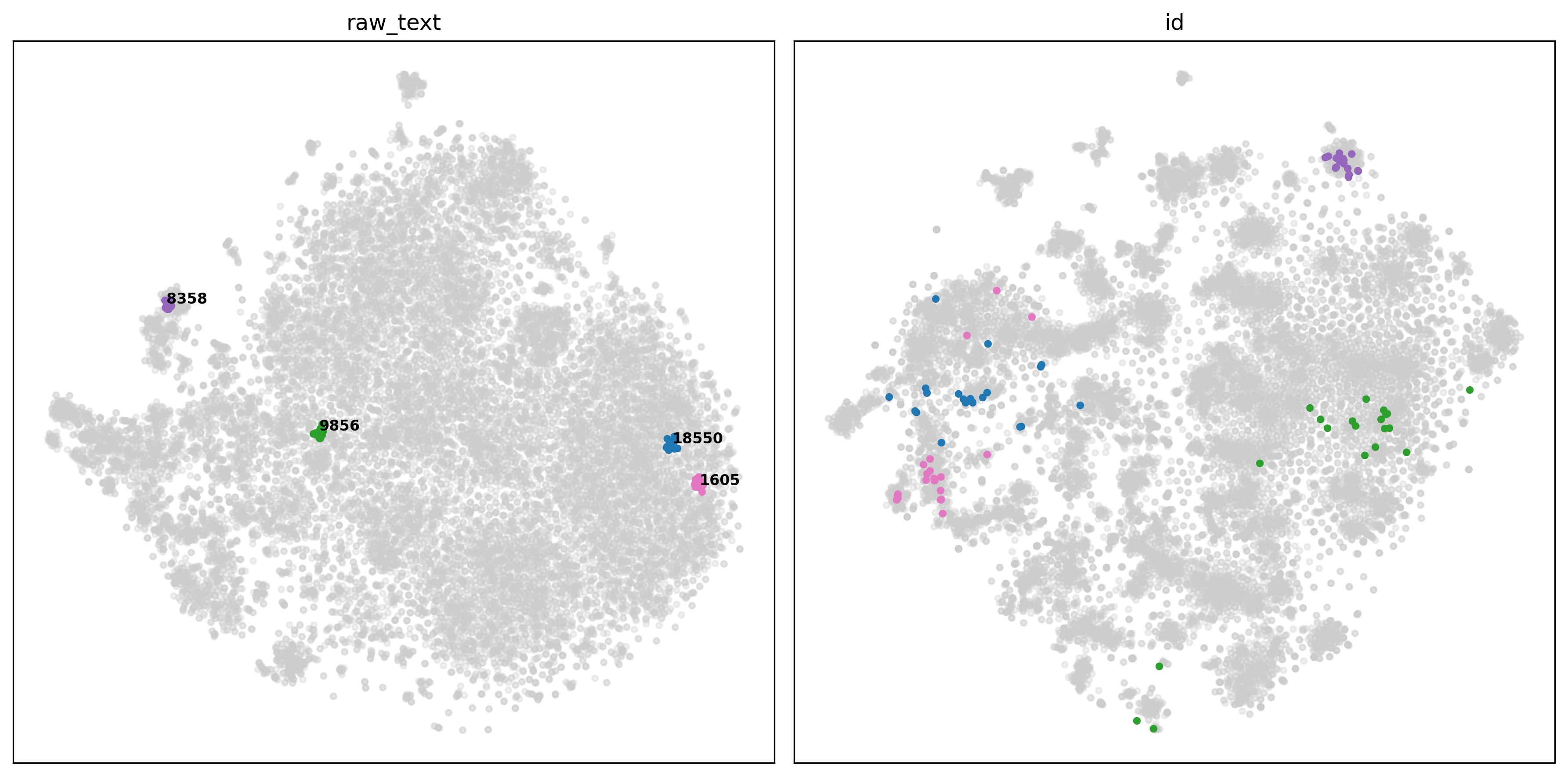}
    \caption{Embedding T-SNE in Books Domain}
    \label{fig:single-emb-tsne}
\end{figure}
To explore the relationship between raw textual semantics and ID embeddings, we visualize four representative item clusters using three embedding views: raw PLM-based text embeddings (without mlp adapter alignment), ID embeddings (Figure.\ref{fig:single-emb-tsne}). As shown in the left subfigure, semantically similar titles (clusters such as MC romance novels, political biographies, kitchen appliance cookbooks, and romantic comedies) naturally form tight clusters in the raw text embedding space. Detail listed in Table \ref{tab:book_clusters}. Notably, the same semantic clusters also appear spatially coherent in the ID embedding space, despite the ID embeddings being trained purely from user interaction signals.

We argue that semantic similarity in textual representations reflects real-world contextual or usage similarity. As a result, items that are semantically close often induce similar interaction patterns, leading to aligned representations in the ID embedding space. 

While AlphaRec\cite{AlphaRec25} demonstrates that textual embeddings can capture collaborative signals through MLP-based processing, our observation goes one step further by revealing that similar patterns emerge directly within ID embeddings. This provides additional evidence for the feasibility of knowledge transfer and alignment between semantic and collaborative spaces, and supports our design where text-driven semantic similarity is used to guide ID embedding migration in cross-domain recommendation.

\begin{table*}
\centering
\caption{Summary of Text Similarity Clusters in the Books Domain}
\renewcommand{\arraystretch}{1.2}
\setlength{\tabcolsep}{4pt}
\begin{tabular}{p{1.5cm} p{2.8cm} p{4.2cm} p{7cm}}
\toprule
\textbf{Cluster ID} & \textbf{Theme} & \textbf{Reader Profile} & \textbf{Representative Titles (Simplified)} \\
\midrule
18550 & MC (Motorcycle Club) Romance & Women 25--45 who enjoy rebellious biker love stories with intense emotion & 
\textit{Ride Rough}, \textit{Cocky Biker}, \textit{Steel (Satan Savages MC Series Book 1) }, \textit{Marked for Death (Blind Jacks MC)}, \textit{Raiden's Choice}, \textit{The Preacher's Daughter (Rough Riders MC)}, \textit{Ride Dirty} \\
\midrule
9856 & Political Biography & Male readers 35+ interested in U.S. history and leadership figures & 
\textit{Alexander Hamilton}, \textit{Grant}, \textit{Truman}, \textit{Destiny of the Republic}, \textit{Washington: A Life}, \textit{Team of Rivals}, \textit{Secret Lives of the First Ladies}, \textit{The Hamilton Affair}, \textit{Hoover}, \textit{The American Spirit} \\
\midrule
8358 & Cooking \& Kitchen & Adults 25--55 seeking fast, healthy home meals using kitchen gadgets & 
\textit{The Easy 5-Ingredient Crock Pot Cookbook}, \textit{Air Fryer Cookbook}, \textit{Instant Pot Cookbook}, \textit{Crock Pot Express Guide}, \textit{Lectin Free Cookbook}, \textit{The Ultimate Cosori Cookbook}, \textit{Healthy Meals for Two}, \textit{Top 500 Instant Pot Recipes}, \textit{Instant Pot for Beginners}, \textit{Simple Air Fryer Recipes} \\
\midrule
1605 & Romantic Comedy & Young women 25--35 who enjoy humorous, sweet, or steamy love stories & 
\textit{Shacking Up}, \textit{Most Valuable Playboy}, \textit{Bought}, \textit{Babyjacked}, \textit{Faking For Him}, \textit{Big Sexy Love}, \textit{Rockstar Retreat}, \textit{Auctioned to the Biker}, \textit{Love Waltzes In}, \textit{Man in Charge} \\
\bottomrule
\end{tabular}

\label{tab:book_clusters}
\end{table*}

\begin{figure}
    \centering
    \includegraphics[width=0.95\linewidth]{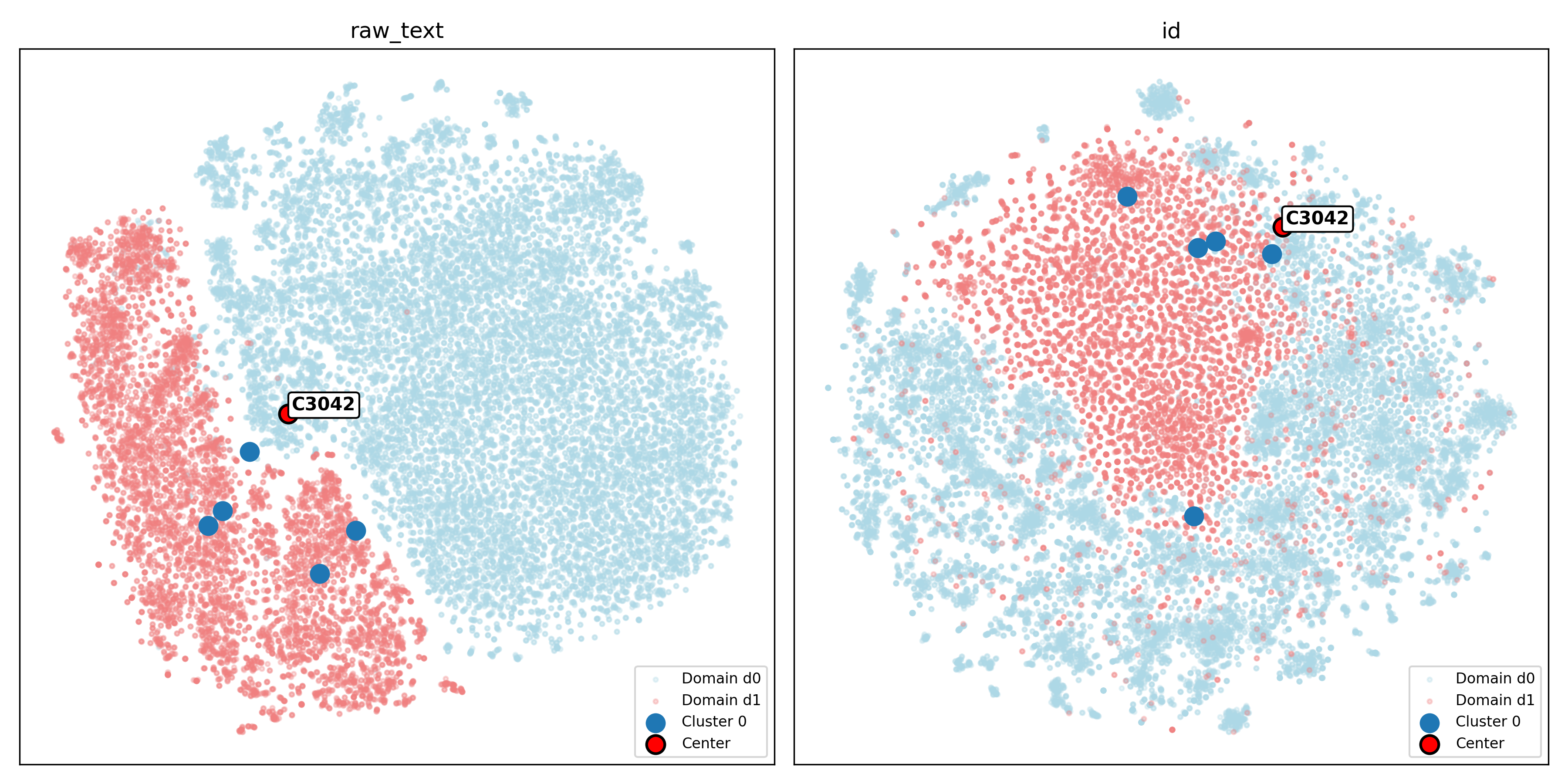}
    \caption{Embedding T-SNE in Books \& Clothes Domain}
    \label{fig:tsn_book_mdr}
\end{figure}
\subsubsection{Cross-Domain Semantic Coherence: Books and Clothes}
To further evaluate the practical effectiveness of cross-domain semantic clustering, we conduct a case study centered on the classic children's book \textit{The Poky Little Puppy}, details listed in Table. \ref{tab:poky_case_study}. As illustrated in Figure.\ref{fig:tsn_book_mdr}, this item in Books Domain forms a coherent cluster in the raw text embedding space with several products from Clothes Domain, namely, children's apparel (e.g., Kids' T-shirts, sweaters, costumes) and children's gifts (e.g., earrings). User reviews across these items exhibit consistent usage scenarios, such as “a gift for my daughter,” “my childhood favorite, now bought for my child,” and “perfect for a baby shower,” all reflecting the common theme of adult-to-child gifting and good emotional bonding. These observations suggest a semantically coherent group characterized by shared usage intention and user type (e.g., parents, grandparents), despite the items belonging to distinct domains.


Further analysis of the ID embedding distributions reveals an interesting contrast. In the trained ID embedding space, which jointly models user-item interactions across domains, we observe that semantically related items maintain close proximity. This implies that shared behavioral patterns, such as purchasing gifts for children, may be implicitly encoded into the learned ID representations.

This finding is consistent with the observations in AlphaRec\cite{AlphaRec25}, which demonstrates that collaborative signals can emerge from text-based representations. More importantly, our result complements this by showing that such semantic commonality may also manifest at the ID embedding level when trained cross-domain graphs. These insights provide further empirical support for our approach, which leverages semantic similarity to guide cross-domain ID migration, and suggest that semantic structure and collaborative behavior are mutually reinforcing in real-world recommendation settings.

\begin{table*}
\centering
\caption{Cross-Domain Case Study Centered on \textit{The Poky Little Puppy}}
\begin{tabular}{@{}p{3.8cm} p{4.5cm} p{8cm}@{}}
\toprule
\textbf{Category} & \textbf{Title} & \textbf{Review Summary} \\
\midrule
\textbf{Central Product (Books)} & \textit{The Poky Little Puppy} & 
“My favorite book as a child, now I bought it for my own child”\newline
“Mom’s favorite book, now gifting it to her”\newline
“Used as a card at a baby shower” \\
\midrule
\textbf{Cross-Domain Neighbors (Clothes)} & \textit{Gildan Long-Sleeve T-Shirt} & 
“Given as a Secret Santa gift for the elderly, also suitable for kids” \\
& \textit{French Toast Girl's Sweater} & 
“Gifted to my 8-year-old granddaughter, fits well and is durable” \\
& \textit{Curious George Hoodie} & 
“Used for a child’s Halloween costume, both cute and practical” \\
& \textit{Moana Girl's Costume} & 
“Sister wore it, and now the younger sister wants one too” \\
& \textit{Sterling Silver Girl's Earrings} & 
“First pair of earrings for my granddaughter, she’s very happy” \\
\bottomrule
\end{tabular}

\label{tab:poky_case_study}
\end{table*}

\subsection{Robustness and Sensitivity Analysis of \modelname{} (RQ6)}
\label{app:robustness}
\subsubsection{Impact of Different LLMs on Model Performance} \label{app:llm_performance}

We conducted experiments to analyze how different Large Language Models (LLMs) affect graph formation and embedding quality in the domains of \textit{Automotive}, \textit{Tools}, \textit{Cell Phones}, and \textit{Clothing $\to$ Sports}. The following table summarizes the key results.

\paragraph{Analysis}

\begin{itemize}
    \item \textbf{Model Size vs. Performance}: Larger models, such as Llama-8B and SFR-Embedding-Mistral-7B, show slight improvements in AUC, Recall, and Precision. However, the performance gains are relatively modest. On the other hand, smaller models like BERT-110M and GPT2-medium-345M still deliver strong performance, providing a good balance between computational efficiency and effectiveness.
    
    \item \textbf{MTEB Benchmark}: We referenced the \textit{MTEB} (Massive Text Embedding Benchmark) , a platform for evaluating text embedding models across tasks like similarity retrieval. to evaluate text embedding models. The SFR-Embedding-Mistral-7B, ranked 5th in the MTEB, achieved the best results in our experiments, highlighting its strengths in semantic indexing and encoding tasks.

    \item \textbf{Sensitivity to LLM Choice}: Our results indicate that the model is not overly sensitive to the choice of LLM. While larger models perform slightly better, smaller models are still sufficient for most applications. We recommend selecting LLMs based on task requirements, available computational resources, and model size.
\end{itemize}

\subsubsection{Sensitivity Analysis of the Threshold $\gamma$} 

\label{app:gamma_sensitivity}

We provide a detailed analysis to support our choice of $\gamma$ and its impact on model performance. Here are our detailed findings:

\begin{table}[h!]
    \centering
    \caption{Impact of different $\gamma$ values on model performance.}
    \begin{tabular}{lccccc}
    \toprule
    \textbf{Metric} & $\gamma = 0.9$ & $\gamma = 0.95$ & $\gamma = 0.99$ & $\gamma = 0.995$ & $\gamma \in [0.6, 0.7]$ \\ 
    \midrule
    AUC             & 0.7490          & 0.7507           & 0.7561           & 0.7511            & 0.7366                 \\ 
    Rec@10 & 0.3249          & 0.3283           & 0.3382           & 0.3359            & 0.3222                 \\ 
    Prec@10 & 0.0548        & 0.0549           & 0.0570           & 0.0561            & 0.0478                 \\ 
    \bottomrule
    \end{tabular}
    
    \label{tab:gamma_performance}
\end{table}

\paragraph{1. Selection of $\gamma$ Based on Similarity Distribution} 

To ensure that the threshold $\gamma$ is both effective and efficient, we conducted a thorough analysis of the similarity distribution among the textual embeddings of items and users across different domains. Specifically, we used the Faiss library to compute the top-20 nearest neighbors for each item/user based on cosine similarity. 

The distribution of these similarities shows that the majority of high-similarity values are concentrated above 0.9. The 5\% quantile is 0.9079, the 25\% quantile is 0.9382, the median (50\% quantile) is 0.9532, the 75\% quantile is 0.9690, and the 95\% quantile is 0.9882. Based on this observation, we initially set the range of $\gamma$ to be 0.9 and above.
\paragraph{2. Extended Experiments with Different $\gamma$ Values}

To further investigate the impact of $\gamma$ on model performance, we conducted experiments with a wider range of $\gamma$ values. We also explored the impact of using lower similarity thresholds (e.g., [0.6, 0.7]) to understand the trade-offs between noise reduction and recall. The results are summarized below:

\paragraph{Analysis of Results}

\begin{itemize}
    \item \textbf{Impact of Higher $\gamma$ Values}: As $\gamma$ increases from 0.9 to 0.99, we observe an improvement in AUC and Mean Recall @10. This indicates that higher $\gamma$ values help in filtering out noise and retaining only the most semantically similar connections, thereby improving the model's ability to transfer knowledge effectively.
    
    \item \textbf{Impact of Very High $\gamma$ Values}: When $\gamma$ is further increased to 0.995, there is a slight drop in AUC and Mean Precision @10. This suggests that overly stringent thresholds may exclude some useful connections, leading to a slight loss in performance.
    
    \item \textbf{Impact of Lower $\gamma$ Values}: Using lower $\gamma$ values (e.g., [0.6, 0.7]) results in a significant drop in performance. This is likely due to the introduction of too many noisy connections, which can degrade the quality of the cross-domain edges and negatively impact the model's ability to learn meaningful representations.
\end{itemize}
\begin{table}[h!]
    \centering
    \caption{Missing data statistics in the 8D dataset from a real-world Amazon business scenario.}
    \begin{tabular}{lcc}
    \toprule
    \textbf{Feature} & \textbf{Missing Ratio} & \textbf{Missing Count} \\ \midrule
    feature          & 0.001711              & 2305                  \\
    salesRank        & 0.003089              & 4161                  \\
    brand            & 0.187505              & 252546                \\ \bottomrule
    \end{tabular}
    
    \label{tab:missing_data}
\end{table}
    
\subsubsection{Generalization to Low-Quality or Noisy Text}

\label{app:low_quality_text}
We have conducted extensive experiments to evaluate the robustness of TextBridgeGNN in domains with low-quality or noisy text. Here are our findings:

\paragraph{1. Real-World Data Analysis}

We used the 8D subset from a real-world Amazon business scenario, which inherently contains sparsity and noise. The statistics of missing data are as follows:

Despite the sparsity and noise, our model achieved an AUC of 0.7561 and Recall@10 of 0.3382, demonstrating its robustness in real-world scenarios.

\paragraph{2. Masking Simulation Experiments}

To further assess robustness, we designed a series of simulation experiments by progressively masking different types of information. The types of information masked include:
\begin{itemize}
    \item ID information (type0),
    \item reviews (type1),
    \item titles (type2),
    \item descriptions, features (type3),
    \item numerical information like price, brand, and salesRank (type4).
\end{itemize}

The results are summarized below:

\begin{table}
\centering
\caption{Results of simulation experiments with different types of masked information.}
\begin{tabular}{lcccc}
\toprule
\textbf{Mask Type} & \textbf{Rate} & \textbf{AUC} & \textbf{Recall@10} & \textbf{Precision@10} \\ \midrule
Ours               & 0                 & 0.7561        & 0.3582              & 0.0570               \\
type0              & 0.1               & 0.7544        & 0.3550              & 0.0524               \\
                   & 0.2               & 0.7508        & 0.3548              & 0.0522               \\
                   & 0.5               & 0.7542        & 0.3604              & 0.0527               \\
type1              & 0.1               & 0.7489        & 0.3471              & 0.0513               \\
                   & 0.2               & 0.7432        & 0.3449              & 0.0508               \\
                   & 0.5               & 0.7261        & 0.3226              & 0.0472               \\
type2              & 0.1               & 0.7548        & 0.3572              & 0.0526               \\
                   & 0.2               & 0.7536        & 0.3569              & 0.0526               \\
                   & 0.5               & 0.7523        & 0.3558              & 0.0523               \\
type3              & 0.1               & 0.7518        & 0.3523              & 0.0521               \\
                   & 0.2               & 0.7524        & 0.3511              & 0.0516               \\
                   & 0.5               & 0.7515        & 0.3529              & 0.0522               \\
type4              & 0.1               & 0.7530        & 0.3541              & 0.0522               \\
                   & 0.2               & 0.7527        & 0.3524              & 0.0520               \\
                   & 0.5               & 0.7507        & 0.3480              & 0.0513               \\
\midrule
Unisrec (full)     & 0                 & 0.6924        & 0.3023              & 0.0521               \\
PEPNet (full)      & 0                 & 0.6967        & 0.3022              & 0.0511               \\
AlphaRec (full)     & 0                 & 0.7031        & 0.3132              & 0.0514               
\\ \bottomrule
\end{tabular}

\label{tab:masking_simulation}
\end{table}

\paragraph{3. Analysis}

\begin{itemize}
    \item \textbf{ID Information}: Removing user and item IDs (type0) has minimal impact on performance, indicating that our model is not overly reliant on ID information.
    
    \item \textbf{Review Information}: Masking reviews (type1) leads to a more noticeable drop in performance, especially when 50\% of reviews are removed. However, even without any review information, our model outperforms baselines like Unisrec and PEPNet, showcasing its robustness in noisy environments.
    
    \item \textbf{Other Textual Information}: Masking descriptions, features, numerical information, and titles (types 2, 3, and 4) results in only minor performance drops, suggesting that these attributes contribute limited value to our model.
\end{itemize}

Overall, while sparse or noisy text can affect performance, our model remains stable and achieves strong results even without certain types of information.

\subsubsection{Adaptability to Cold-Start Domain-Adaptation Tasks} \label{app:coldstart}
\vspace{-6pt}
\begin{table}[h!]
    \centering
    \caption{Cold-start results on the Sports domain}
    \begin{tabular}{lccc}
    \toprule
    \textbf{Method}                    & \textbf{AUC} & \textbf{Recall@10} & \textbf{Precision@10} \\ 
    \midrule
    UniSRec (fully trained)           & 0.5328       & 0.0787             & 0.0103                \\
    EDDA (fully trained)    & 0.5215       & 0.1034             & 0.0151                \\
    LightGCN (fully trained)          & 0.5022       & 0.0820             & 0.0114                \\
    AlphaRec (fully trained) & \underline{0.5591} & \underline{0.1211} & \underline{0.0162}        \\
    Ours (fully trained)    & \textbf{0.5723}       & \textbf{0.1379}             & \textbf{0.0190}                \\
    \textit{Relative improvement}     & \textit{+2.36\%} & \textit{+13.87\%} & \textit{+17.28\%} \\
    \midrule
    AlphaRec (zeroshot)      & \underline{0.5220} & \underline{0.1178} & \underline{0.0162}        \\
    Ours (zeroshot)         & \textbf{0.5424}       & \textbf{0.1256}             & \textbf{0.0168}                \\
    \textit{Relative improvement}     & \textit{+3.91\%} & \textit{+6.62\%}  & \textit{+3.70\%} \\
    \bottomrule
    \end{tabular}
    \label{tab:coldstart_results}
    \end{table}

We have explored the adaptability of our model to cold-start and domain-adaptation tasks beyond recommendation. Here are our findings:

\paragraph{1. Cold-Start Experiment}

We simulated a cold-start scenario by retaining only 5\% of the original training data in the target domain (Sports) while keeping other data unchanged. The results are as follows:

\paragraph{2. Analysis}

\begin{itemize}
    \item \textbf{Sparse Data}: The low performance of most models is primarily due to the extreme sparsity of training data in the target domain. This makes it challenging for models to generalize and perform well with minimal training data.
    
    \item \textbf{Cold-Start Capability}: Our model significantly outperforms other baselines, demonstrating its ability to adapt to cold-start scenarios. Even in a training-free setting, our model achieves strong results, highlighting its robustness and adaptability to limited data availability.
\end{itemize}

\subsection{Complexity Analysis}  \label{app:complexity}

Assume there are $n$ domains (with $n-1$ source domains and one target domain), each with approximately $N$ nodes and $E$ edges, and that additional semantic edges $E_{sem}$ are constructed across domains. The original feature dimension is $2H$, with a Text Adapter hidden size of $H$, and each node requires an LLM call with cost $t_{\text{LLM}}$. 

In data preprocessing, generating text embeddings for all $nN$ nodes incurs a cost of $O(nN\,t_{\text{LLM}})$, and constructing similarity edges without acceleration would cost $O(H\cdot n(n-1)N^2)$. However, using Faiss, each node retrieves a constant number of nearest neighbors, reducing computation cost to $O(nN\,H)$. Hence, the overall preprocessing complexity is $O(nN\,t_{\text{LLM}} + nN\,H)$.

During training or inference, source domain pre-training (including local propagation over $E$ edges and Text Adapter updates on $N$ nodes per domain, plus global propagation over $(n-1)N$ nodes with $(n-1)E+E_{sem}$ edges) incurs a cost of $O(nE\,H + E_{sem}\,H + nN\,H^2)$, and target domain adaptation contributes approximately $O(nE\,H + E_{sem}\,H + N\,H^2)$. Thus, the total training cost is $O(nE\,H + E_{sem}\,H + nN\,H^2)$. If $E$ and $N$ are considered as primary variables, the complexity simplifies to:
$O(E + N)$

\vspace{0.5em}
\noindent
\textbf{Efficiency.} We provide an cost-effectiveness comparison on a single RTX 3090 24GB GPU using the 8D dataset:

\begin{itemize}
    \item \textbf{Memory Usage:} \modelname{} requires approximately 12GB of GPU memory, whereas UniSRec consumes around 20GB.
    \item \textbf{Training Time:} Each epoch of our model takes roughly 1 minute, while UniSRec requires around 6 minutes per epoch in both pre-training and fine-tuning stages.
\end{itemize}

These results indicate that \modelname{} maintains a favorable balance between effectiveness and efficiency, avoiding excessive memory or time overhead compared to baselines.
\begin{table*}[t!]
\centering
\caption{Performance of different LLMs on the Automotive, Tools, Cell Phones, Clothing $\to$ Sports domains.}
\begin{tabular}{lccccc}
\toprule
\textbf{Model}                    & \textbf{AUC}  & \textbf{Recall@10} & \textbf{Recall@20} & \textbf{Precision@10} & \textbf{Precision@20} \\ \midrule
BERT-110M                         & 0.7273        & 0.332             & 0.469             & 0.0443                & 0.0311               \\
GPT2-medium-345M                  & 0.7408        & 0.3418            & 0.4749            & 0.0506                & 0.0354               \\
Llama-8B                          & 0.7506        & 0.3618            & 0.4981            & 0.0522                & 0.0365               \\
SFR-Embedding-Mistral-7B          & 0.7579        & 0.3644            & 0.5036            & 0.0524                & 0.037                \\ \bottomrule
\end{tabular}

\label{tab:llm_performance}
\end{table*}

\subsection{Fusion Strategy Comparison}
\label{app:fusion}

To evaluate whether the choice of fusion strategy affects the framework, we compare three alternatives under identical hyperparameters on the 8D$\to$Sports dataset:

\begin{table}[H]
\centering
\caption{Comparison of fusion strategies (8D$\to$Sports).}
\begin{tabular}{lcc}
\toprule
\textbf{Fusion Strategy} & \textbf{AUC} & \textbf{Recall@10} \\ \midrule
Addition (Ours)          & 0.7506       & 0.3618            \\
Gated Fusion             & 0.7584       & 0.3680            \\
Dim-wise Cross-Attention & 0.7560       & 0.3716            \\ \bottomrule
\end{tabular}
\label{tab:fusion_comparison}
\end{table}

\noindent
\textbf{Gated Fusion} applies a per-dimension learnable gate. \textbf{Dim-wise Cross-Attention} splits the $d$-dimensional embedding into 4 groups and applies cross-attention across them.

Complex fusion methods show marginal advantages on certain metrics (e.g., +0.78\% AUC for gated fusion). However, simple addition achieves competitive results (AUC 0.7506), confirming that the hierarchical graph propagation and ID transfer mechanism (the core contribution) effectively captures ID-text interactions regardless of fusion choice. Exploring adaptive fusion integrated into GNN propagation layers is a promising direction for future work.

\subsection{Failure Case Analysis: Semantic vs.\ Behavioral Similarity}
\label{app:failure_case}

To investigate cases where semantic similarity conflicts with behavioral patterns, we track cosine similarity in two spaces: text embeddings (from the LLM) and model embeddings. A representative failure case is a work boot (Tools \& Home Improvement) vs.\ a sports boot (Sports \& Outdoors)---both describe similar boot products, yielding a text embedding cosine of 0.96, but they barely have any behavioral overlap. This leads to a model embedding cosine of 0.314 in the zero-shot setting and $-0.07$ after fine-tuning. Despite the high text cosine, the model cosine is only moderate at zero-shot and further separates after fine-tuning, showing that text similarity acts as a \textit{soft prior} that is corrected by collaborative signals.

\subsection{End-to-End Computational Cost}
\label{app:end2end_cost}

We provide a detailed breakdown of the computational cost on an RTX 3090 (24GB) for the 8D$\to$Sports transfer task. Note that per-epoch time is the fairest metric for efficiency comparison, as total training time depends on convergence speed which varies with hyperparameters and model design. We report both for completeness:

\begin{table}[H]
\centering
\caption{End-to-end training time comparison (8D$\to$Sports, RTX 3090).}
\begin{tabular}{lcc}
\toprule
\textbf{Method} & \textbf{Per-Epoch} & \textbf{Total Time} \\ \midrule
\modelname{}     & ~65s    & ~34 min  \\
EDDA             & ~63s    & ~32 min  \\
LightGCN         & ~75s    & ~39 min  \\
AlphaRec         & -       & ~1.5 h   \\
UniSRec          & ~6 min  & ~15.3 h  \\ \bottomrule
\end{tabular}
\label{tab:end2end_cost}
\end{table}

\noindent
\textbf{Semantic Graph Construction.} Faiss-gpu (IVF+PQ index) constructs the similarity graph for 107K items in ~40 seconds, reducing complexity from $O(N^2)$ to $O(N \cdot K)$ where $K$ is the number of nearest neighbors.

\noindent
\textbf{Dynamic Updates.} When new items arrive, only the new item's embedding and Faiss neighbor search are needed, followed by incremental edge insertion; no full graph rebuild is required.

\noindent
\textbf{LLM Embedding.} The LLM is used offline as a one-time feature extraction step and does not affect training or inference efficiency. The framework is agnostic to the embedding model.

\subsection{Cross-Platform Transfer}
\label{app:cross_platform}

To validate generalizability beyond same-platform cross-category transfer, we conduct a cross-platform experiment: pretrain on Amazon (8D collection) and finetune on Steam Action games. We use the same time-interval and core-10 filtering as in our main experiments, resulting in 56K interactions. Despite the significant platform gap (e-commerce vs.\ gaming), \modelname{} achieves +2.3\% AUC and +11.1\% Recall@10 over LightGCN, suggesting the framework's transfer mechanism is applicable beyond Amazon-style recommendation settings.

\subsection{Diversity and Coverage}
\label{app:beyond_accuracy}

We also examine whether the accuracy gains affect recommendation diversity on 8D$\to$Sports. ILD@20 is computed from item-category similarity, and Coverage@20 measures the fraction of the target catalog covered by the top-20 lists. As shown in Table.\ref{tab:beyond_accuracy}, \modelname{} maintains comparable diversity (ILD@20 0.755 vs.\ 0.750) and slightly higher coverage (0.151 vs.\ 0.142) than LightGCN; coverage also grows from 0.145 (training-free) to 0.151 after fine-tuning, suggesting that fine-tuning activates additional long-tail items rather than concentrating on popular ones.

\begin{table}[H]
\centering
\caption{Diversity and coverage on 8D$\to$Sports.}
\begin{tabular}{lcc}
\toprule
\textbf{Method} & \textbf{ILD@20} & \textbf{Coverage@20} \\ \midrule
LightGCN        & 0.750 & 0.142 \\
\modelname{}    & 0.755 & 0.151 \\ \bottomrule
\end{tabular}
\label{tab:beyond_accuracy}
\end{table}

\end{document}